\documentclass[12pt]{iopart}

\usepackage{iopams}
\usepackage{graphicx,colordvi}
\usepackage{amssymb,slashed,cite}
\expandafter\let\csname equation*\endcsname\relax
\expandafter\let\csname endequation*\endcsname\relax
\usepackage{amsmath}

\usepackage{booktabs,tabulary}
\usepackage{dsfont}
\usepackage{color}
\usepackage[breaklinks=true,colorlinks=true,linkcolor=blue,urlcolor=blue,citecolor=blue]{hyperref}
\usepackage[gen]{eurosym}

\newcommand{\GeV}{\,\text{GeV}}

\newcommand{\toright}[1]{\hspace*{\fill}{\footnotesize{#1}}}

\begin{document}

\title[Baryon Number Violation: From Nuclear Matrix Elements to BSM Physics]{\toright{\textnormal{INT-PUB-25-009, LA-UR-25-23551, PSI-PR-25-07, YITP-SB-2025-08, ZU-TH 23/25}}\\[0.5cm]Baryon Number Violation: From Nuclear Matrix Elements to BSM Physics}

\author{Leah J.\ Broussard$^1$, Andreas Crivellin$^2$, Martin Hoferichter$^3$\footnote{hoferichter@itp.unibe.ch}, Sergey Syritsyn$^4$, Yasumichi Aoki$^5$, Joshua L.~Barrow$^6$, Arnau Bas i Beneito$^{7,8}$,  	Zurab Berezhiani$^{9,10}$,  Nicola Fulvio Calabria$^{11}$,  	Svjetlana Fajfer$^{12,13}$, Susan Gardner$^{14}$,  	Julian Heeck$^{15}$, 	Cailian Jiang$^{16}$, Luca Naterop$^{2,17}$, Alexey A.~Petrov$^{18}$, Robert Shrock$^{19}$, Adrian Thompson$^{20}$,  Ubirajara van Kolck$^{21,22,23}$, Michael L.~Wagman$^{24}$, Linyan Wan$^{24}$, John Womersley$^{25}$, Jun-Sik Yoo$^{26}$}

\address{$^1$ Oak Ridge National Laboratory, Oak Ridge, TN 37831, USA
}

\address{$^2$ Physik-Institut, Universit\"at Z\"urich, Winterthurerstrasse 190, 8057 Z\"urich, Switzerland}

\address{$^3$ Albert Einstein Center for Fundamental Physics, Institute for Theoretical Physics, University of Bern, Sidlerstrasse 5, 3012 Bern, Switzerland}

\address{$^4$ Center for Nuclear Theory, Department of Physics and Astronomy, Stony Brook University, NY 11794-3800, USA}

\address{$^5$ RIKEN Center for Computational Science, Kobe 650-0047, Japan}

\address{$^6$ The University of Minnesota, Minneapolis, MN 55455, USA}

\address{$^{7}$ Departament de F\'isica Te\`orica, Universitat de Val\`encia, 46100 Burjassot, Spain}
\address{$^{8}$ Institut de F\'isica Corpuscular (CSIC-Universitat de Val\`encia), Parc Cient\'ific UV, C/Catedr\'atico Jos\'e Beltr\'an, 2, 46980 Paterna, Spain}

\address{$^{9}$ Dipartimento di Fisica e Chimica, Universit\`a
di L'Aquila, 67010 Coppito AQ, Italy}

\address{$^{10}$ INFN, Laboratori Nazionali del Gran Sasso, 67010 Assergi AQ, Italy}

\address{$^{11}$ INFN Sezione di Bari and Politecnico di Bari, Dipartimento Interuniversitario di Fisica, Bari, Italy}

\address{$^{12}$ Department of Physics, University of Ljubljana, Jadranska 19, 1000 Ljubljana, Slovenia}
\address{$^{13}$
Jo\v zef Stefan Institute, Jamova 39, 1000  Ljubljana, Slovenia}

\address{$^{14}$ Department of Physics and Astronomy, University of Kentucky, Lexington, KY 40506-0055, USA}

\address{$^{15}$ Department of Physics, University of Virginia,
Charlottesville, Virginia 22904, USA}

\address{$^{16}$ Nanjing University, Nanjing, China}

\address{$^{17}$ Paul Scherrer Institut, 5232 Villigen PSI, Switzerland}

\address{$^{18}$ Department of Physics and Astronomy,
University of South Carolina,
Columbia, SC 29208, USA}

\address{$^{19}$ C.~N.~Yang Institute for Theoretical Physics and Department of Physics and
Astronomy, Stony Brook University, Stony Brook, NY 11794, USA}

\address{$^{20}$ Northwestern~University,~Evanston,~IL~60208,~USA}

\address{$^{21}$ European Centre for Theoretical Studies in Nuclear Physics and Related Areas (ECT*), Fondazione Bruno Kessler, 38123 Villazzano Trento, Italy}

\address{$^{22}$ Universit\'e Paris-Saclay, CNRS/IN2P3, IJCLab, 91405 Orsay, France}

\address{$^{23}$ Department of Physics, University of Arizona, Tucson, AZ 87521, USA}

\address{$^{24}$ Fermi National Accelerator Laboratory, Batavia, IL 60510, USA}

\address{$^{25}$ University of Edinburgh, United Kingdom}

\address{$^{26}$ Los Alamos National Laboratory, Theoretical Division T-2, Los Alamos, New Mexico 87545, USA}

\begin{abstract}
Processes that violate baryon number, most notably proton decay and $n\bar n$ transitions, are promising probes of physics beyond the Standard Model (BSM) needed to understand the lack of antimatter in the Universe. To interpret current and forthcoming experimental limits, theory input from nuclear matrix elements to UV complete models enters. Thus, an interplay of experiment, effective field theory, lattice QCD, and BSM model building is required to develop strategies to accurately extract information from current and future data and maximize the impact and sensitivity of next-generation experiments. Here,
we briefly summarize the main results and discussions from the workshop ``INT-25-91W: Baryon Number Violation: From Nuclear Matrix Elements to BSM Physics,'' held at the Institute for Nuclear Theory, University of Washington, Seattle, WA, January 13--17, 2025.
\end{abstract}

\newpage

\tableofcontents

\markboth{Baryon Number Violation: From Nuclear Matrix Elements to BSM Physics}{Baryon Number Violation: From Nuclear Matrix Elements to BSM Physics}

\newpage

\section{Introduction}

Testing the possible violation of symmetries is a very promising avenue in the search for physics beyond the Standard Model (BSM). In this context, baryon number violation (BNV)
is particularly interesting, probing the highest scales of all BSM searches, exceeding $10^{15}\GeV$ in the most sensitive search channels in proton decay. This, therefore, sets the most stringent limits on the BSM scales of the associated higher-dimensional operators and the viability of grand-unified-theory (GUT) scenarios~\cite{Pati:1973uk,Georgi:1974sy,Fritzsch:1974nn}.

Further, BNV is a key prerequisite for baryogenesis via Sakharov's conditions~\cite{Sakharov:1967dj}. Baryon number ($B$) needs to be violated to produce the observed baryon asymmetry in the Universe, yet still has to be discovered experimentally. Efforts to directly observe BNV are ongoing in searches for nucleon decays (in particular proton decay)---in the future further pushing the sensitivity with Hyper-Kamiokande~\cite{Hyper-Kamiokande:2018ofw}, DUNE~\cite{DUNE:2020ypp}, and JUNO~\cite{JUNO:2021vlw}---and planned in searches for neutron--antineutron ($n\bar n$) transitions with NNBAR~\cite{Backman:2022szk}. A selection of current limits as listed by the Particle Data Group (PDG)~\cite{ParticleDataGroup:2024cfk} is shown in Table~\ref{tab:lifetime}. These include flagship processes such as $p\to\pi^0\ell^+$, $\ell=e,\mu$~\cite{Super-Kamiokande:2020wjk}, $p\to K^+\bar\nu$~\cite{Super-Kamiokande:2014otb}, and other single-meson final states~\cite{Super-Kamiokande:2013rwg,Super-Kamiokande:2005lev,Super-Kamiokande:2022egr,Super-Kamiokande:2017gev,McGrew:1999nd}, but also vector-meson and radiative final states~\cite{Super-Kamiokande:2017gev,McGrew:1999nd} as well as inclusive limits~\cite{Learned:1979gp,Cherry:1981uq}.  Best bound and free limits on $n\bar n$ transitions are also given~\cite{Super-Kamiokande:2020bov, Baldo-Ceolin:1994hzw}.

\begin{table}[t]
	\renewcommand{\arraystretch}{1.3}
	\centering
	\scalebox{0.9}{
	\begin{tabular}{l r r r}
		\toprule
		 Channel & Limit [$10^{30}$ y] & & Reference \\\midrule
		 $p\to \pi^0 e^+$ & $2.4\times 10^{4}$ & & \cite{Super-Kamiokande:2020wjk}\\
		 $p\to \pi^0 \mu^+$ & $1.6\times 10^{4}$ & & \cite{Super-Kamiokande:2020wjk}\\
		 $p\to \pi^+ \bar \nu$ & $3.9\times 10^{2}$ & & \cite{Super-Kamiokande:2013rwg}\\
		 $p\to K^0e^+$& $1.0\times 10^3$& & \cite{Super-Kamiokande:2005lev}\\
		 $p\to K^0\mu^+$& $3.6\times 10^3$& & \cite{Super-Kamiokande:2022egr}\\
		 $p\to K^+\bar\nu$& $5.9\times 10^3$& & \cite{Super-Kamiokande:2014otb}\\
		  $p\to \eta e^+$& $1.0\times 10^4$& & \cite{Super-Kamiokande:2017gev}\\
		 $p\to\eta\mu^+$& $4.7\times 10^3$& & \cite{Super-Kamiokande:2017gev}\\\midrule
		 $n\to \pi^- e^+$ & $5.3\times 10^3$ & & \cite{Super-Kamiokande:2017gev}\\
		 $n\to \pi^- \mu^+$ & $3.5\times 10^3$ & & \cite{Super-Kamiokande:2017gev}\\
		 $n\to \pi^0 \bar \nu$ & $1.1\times 10^{3}$ & & \cite{Super-Kamiokande:2013rwg}\\
		 $n\to K^0\bar\nu$& $1.3\times 10^2$& & \cite{Super-Kamiokande:2005lev}\\
		 $n\to \eta\bar\nu$& $1.6\times 10^2$& & \cite{McGrew:1999nd}\\\midrule
$p\to \rho^0e^+$ & $7.2\times 10^2$ & & \cite{Super-Kamiokande:2017gev}\\
$p\to \rho^0\mu^+$ & $5.7\times 10^2$ & & \cite{Super-Kamiokande:2017gev}\\
$p\to \omega e^+$ & $1.6\times 10^3$ & & \cite{Super-Kamiokande:2017gev}\\
$p\to \omega\mu^+$ & $2.8\times 10^3$ & & \cite{Super-Kamiokande:2017gev}\\
$p\to K^{*,0}e^+$ & $84$ & & \cite{McGrew:1999nd}\\
$p\to K^{*,+}\bar \nu$ & $51$ & & \cite{McGrew:1999nd}\\
		 $p\to e^+\gamma$ & $6.7\times 10^2$ & & \cite{McGrew:1999nd}\\
		 $p\to \mu^+\gamma$ & $4.8\times 10^2$ & & \cite{McGrew:1999nd}
		 \\\midrule
		 $p,n\to e^+ X$ & $0.6$ & & \cite{Learned:1979gp}\\
		 $p,n\to \mu^+ X$ & $12$ & & \cite{Cherry:1981uq}\\
		 \midrule
		 Channel & Limit [$10^{30}$ y] & $\tau_{n\bar{n}}$ [$10^{8}$ s] & Reference \\
		 \midrule
		 bound $n\rightarrow\bar{n}$ & $3.6\times 10^2$ & $4.7$ & \cite{Super-Kamiokande:2020bov}\\
		 free $n\rightarrow\bar{n}$ & & $0.86$  & \cite{Baldo-Ceolin:1994hzw}\\
		\bottomrule
	\end{tabular}
	}
	\caption{Limits on nucleon lifetimes for various decay channels studied in water-Cherenkov detectors and on $n\bar n$ oscillation times, with equivalent free oscillation times $\tau_{n\bar{n}}$, all at $90\%$ confidence level (C.L.)~\cite{ParticleDataGroup:2024cfk}. }
	\label{tab:lifetime}
\end{table}

Digesting these experimental capabilities has far-reaching implications. A positive signal would provide unambiguous direct evidence of the breakdown of a key prediction of the SM and might even hint at some low-scale baryogenesis mechanism. Just as importantly, non-observation provokes many 
questions for the fundamental physics community. How should we contextualize these
measurements when rendered, and what does this generally mean for theories of baryogenesis? Would it be
possible for various models of proton decay and $n\bar n$ transitions to be
differentiated if either signal were observed? If not, what constraints on models are possible? Can certain $B$-violating models be eliminated? What does this imply for other hypothetical physical mechanisms operating at higher scales, e.g., GUT-scale baryogenesis and Seesaw models?

From the perspective of nuclear theory, these questions are intimately related to the calculation of nuclear matrix elements, though  
single-nucleon matrix elements or nuclear-matter effects may (also) require assessment. 
At the high scale, BNV can be parameterized via effective operators in SM effective field theory (SMEFT), but the observables are defined at the low, nuclear scale where the experiments are performed. For a robust calculation of these observables in terms of the underlying Wilson coefficients, one therefore needs both renormalization group evolution (RGE) and, crucially, the hadronization of the effective operators. During the  workshop  ``INT-25-91W: Baryon Number Violation: From Nuclear Matrix Elements to BSM Physics,'' held at the Institute for Nuclear Theory, University of Washington,  Seattle, WA, January 13--17, 2025~\cite{INT}, these interrelated questions were discussed. In particular, the workshop addressed 
how to extract the maximum amount of information from the available data from current and future experiments and determine their impact on the BSM landscape. In Secs.~\ref{sec:exp}--\ref{sec:BSM} we report on some of the main discussion outcomes, followed by short summaries of frontier developments
in Sec.~\ref{sec:abstracts}, reflecting perspectives from experiment, EFT, lattice QCD (LQCD), and BSM model building.

\section{Experiments}
\label{sec:exp}

The workshop opened with presentations from ongoing and future experiments probing BNV including proton decay at Super-/Hyper-Kamiokande, DUNE, and JUNO (see Secs.~\ref{sec:HyperK}, \ref{sec:DUNE}, and \ref {sec:JUNO}), as well as $n\bar n$ transitions at DUNE, JUNO, and HIBEAM/NNBAR (see Secs.~\ref{sec:nnbar}, \ref {sec:JUNO}, and \ref{sec:NNBAR}). These presentations initiated detailed discussions about analysis strategies, sensitivities, and theory input for the interpretation of measurements, e.g., related to the modeling of the nuclear interaction in proton-decay searches and the conversion of $n\bar n$ transitions in a nucleus to the free-neutron $n\bar n$ transition rate. There was strong interest from the theory side to also consider channels besides the golden modes $p\to e^+\pi^0$ and $p\to \nu K^+$ to close other flat regions in EFT parameter space. An important revelation was that the Super-Kamiokande lifetime limits for $p\to e^+\gamma$ ($4.1\times 10^{34}$ y) and $p\to\mu^+\gamma$ ($2.1\times 10^{34}$ y) from Ref.~\cite{Super-Kamiokande:2018apg} are not yet published. These limits currently do not feature in the PDG's review~\cite{ParticleDataGroup:2024cfk}, despite being the
largest lifetime limits ever reported, exceeding the current official limits by two orders of magnitude. Accordingly, the analysis should be pushed forward to publication.
In view of possible resonance enhancement of the matrix elements, the feasibility of searches with vector-meson final states was also discussed, see Sec.~\ref{sec:EFT+lattice}.

In addition, to capture a multitude of many-body decay channels, it was proposed to conduct more inclusive searches.
These inclusive searches will be less sensitive to specific channels than corresponding exclusive searches, due to the exclusive searches utilizing final-state topologic and kinetic constraints to reject atmospheric neutrino backgrounds.
Reporting inclusive searches in energy differential bins would mitigate this disadvantage.
This approach allows the experiment to recover part of the background rejection handle in kinetic constraints and facilitates easier application to specific theory-motivated channels while maintaining the nature of being inclusive.
For example, in the case of $p\to e^++\text{anything}$, instead of assuming two-body invisible decay~\cite{Super-Kamiokande:2015pys} or certain final states (e.g., Refs.~\cite{Super-Kamiokande:2020tor,Super-Kamiokande:2020wjk,Super-Kamiokande:2024qbv,Super-Kamiokande:2017gev}), we propose to first construct an inclusive sample with at least one $e^+$ in the events regardless of other features, and report the $e^+$ energy spectrum as well as the conversion into proton lifetime constraint assuming a mono-energetic signature.
The uncertainty budget in inclusive searches is typically dominated by statistical uncertainty, due to the worse background rejection and the absence of the usually dominant uncertainty when modeling nuclear effects in proton decay.
When applying the inclusive search limit to an exotic channel with $e^+$ in the final state that does not have a dedicated exclusive search, for example, $p\to e^+\pi^0\nu\nu$, one will first calculate the probability density spectrum of the $e^+$ in $e^+$ energy, then convolve it with the inclusive search result of $p\to e^++\text{anything}$ differential in $e^+$ energy.
Nuclear uncertainty should be applied to the probability density spectrum estimation when feasible.
We envision in the future, exclusive searches will provide proton lifetime limits for theoretically well-motivated channels, while such kind of differential inclusive searches provide a wide coverage of proton decay and BNV processes.

In the context of $n\bar n$ and $p\bar n$ searches, it was observed that the inference of expected branching fractions from isospin symmetry~\cite{Super-Kamiokande:2020bov} via $p\bar p$~\cite{CrystalBarrel:2003uej,Klempt:2005pp} and $n\bar p$~\cite{Bressani:2003pv} experiments could be improved by decomposing the various final states into isospin components and thereby disentangling the isospin-$0$ and -$1$ contributions, see Ref.~\cite{Ditsche:2012fv} for the $N\bar N\to2\pi$ case ($N\in\{p,n\}$). The observation that $(4\text{--}6)$-pion states have the largest branching fraction can be explained by analogy to hadronic vacuum polarization $e^+e^-\to\text{hadrons}$~\cite{Aoyama:2020ynm}. In this case, at $e^+e^-$ center-of-mass energies that correspond to $N\bar N$, multi-pion channels dominate over, e.g., the two-pion mode, since at these higher energies, resonances couple increasingly strongly to high-multiplicity channels and the threshold suppression fades away.

\section{Effective field theory and lattice QCD}
\label{sec:EFT+lattice}

To make the connection between the low-energy scale at which the experimental searches proceed and the high-energy scale at which the BNV physics occurs, there were two presentations featuring SMEFT. A comprehensive one-loop analysis, see Sec.~\ref{sec:SMEFT}, showed that some channels were critical in resolving flat directions in parameter space, most notably the $p\to e^+\eta$ channel, due to different flavor structure compared to the golden mode $p\to e^+\pi^0$~\cite{Beneito:2023xbk}. For the $\eta$ channel, no direct LQCD calculations of the
proton matrix elements are currently available due to the complications associated with disconnected diagrams~\cite{Yoo:2021gql}. The relevance for the SMEFT program motivates such calculations. There was also a presentation on the ongoing program to extend the RGE analysis to two-loop order~\cite{Naterop:2024ydo} (Sec.~\ref{sec:2loop}), since the large separation of scales suggests that the sensitivity of BNV processes to two-loop corrections should be enhanced. However, it was emphasized that a consistent analysis requires the calculation of one-loop matching corrections at both the high-energy scale and in the nuclear matrix elements, i.e., even radiative corrections in the latter if QED running is to be included.

In the $n\bar n$ context, there was a presentation highlighting the intricacies of the calculation of the nuclear conversion factor for the $n\bar n$ transition in deuterium, especially the sensitivity to the power counting in the nuclear EFT~\cite{Oosterhof:2019dlo}, see Sec.~\ref{sec:EFT_nuclear}. Currently, the precision is limited by two-nucleon contact terms, which could conceivably be determined from LQCD, as an application of the ongoing spectroscopy program in the $NN$ sector, see Sec.~\ref{sec:lattice_nnbar}. In this way, the calculation for deuterium could be improved, the power counting tested, and better few-nucleon amplitudes provided for nuclear-structure calculations in heavier nuclei. In contrast, two- and more-nucleon effects can, to a very good approximation, be neglected for proton decay~\cite{Oosterhof:2021lvt}.

Regarding proton-decay matrix elements (see Secs.~\ref{sec:lattice_proton_decay} and~\ref{sec:lattice_proton_decay_2}), the feasibility of performing LQCD calculations for resonance final states such as $p\to \rho^0 \ell^+$ or $p\to K^{*,+}\nu$ was discussed, given the potential extended sensitivity in case the matrix elements for these states indeed display some resonance enhancements. Critically, given the substantial theoretical effort, better projections for the expected experimental sensitivity were called for, and the $p\to K^{*,+}\nu$ channel was identified as one case in which limits could become competitive. In this regard, a potentially interesting connection to $p\to \ell^+\gamma$ was identified, in which vector-meson-dominance arguments could potentially be used to relate the more complicated resonance matrix elements to photon ones to get a first estimate.

Finally, EFT methods can also be used to establish connections to the third generation~\cite{Beneke:2024hox}, including BNV $B$-decays and proton decays involving an off-shell $\tau$ (Sec.~\ref{sec:third_gen}). In this case, flat directions in the EFT can still be closed, but the BNV scale can be lower. On the theory side, it would be interesting to validate factorization assumptions necessary for some matrix elements with LQCD calculations, while the momentum dependence of BNV form factors can also be reconstructed using dispersion relations~\cite{Crivellin:2023ter}. Moreover, in case of third-generation-induced nucleon decays, higher-multiplicity final states occur due to the decays of the $\tau$ lepton, further motivating the consideration of inclusive searches discussed in Sec.~\ref{sec:exp}.

\section{BSM and astrophysics connections}
\label{sec:BSM}

The landscape of possible, distinct BNV scenarios is vast, and can be organized, for instance, by $B\pm L$ and $L_\ell$, $\ell=e,\mu,\tau$ quantum numbers~\cite{Heeck:2024jei} ($L$ referring to total lepton number and $L_\ell$ that of a given flavor). Interestingly, even for higher-dimensional operators BSM realizations exist that avoid at the same time contributions at lower dimensions, see Sec.~\ref{sec:opening}. To fully probe the BNV landscape, one would need to measure an unrealistically large number of channels, including high-multiplicity final states. This motivates the consideration of inclusive limits, ideally including information on the spectrum, together with a few theory-motivated exclusive modes. Besides the golden modes $p\to \pi^0\ell^+$ and $p\to K^+\nu$, these could include $p\to \eta\ell^+$ to break degeneracies in flavor space, $p\to\rho\ell^+$ or $p\to K^{*,+}\nu$ due to the potential resonance enhancement, and $p\to \ell^+\gamma$ due to large sensitivity reach. In the latter case, however, estimates of the matrix elements do suggest a suppression by the fine-structure constant $\alpha$, but refined calculations especially in view of the (unpublished) limit from Ref.~\cite{Super-Kamiokande:2018apg} appear well motivated.

As specific BSM realizations, leptoquark models were discussed (Sec.~\ref{sec:leptoquark})---in which proton decay can be absent at tree level but induced by loop effects---as were scenarios in
which $n\bar n$ transitions would be the dominant manifestation of BNV (Sec.~\ref{sec:nnbar_theory}). Finally, connections
to astrophysics in the context of BNV and neutron disappearance to dark-sector final
states were explored, together with possible connections to dark matter, late-scale
baryogenesis models, and the tension in the neutron lifetime between beam and bottle measurements~\cite{Czarnecki:2018okw}, see Secs.~\ref{sec:cosmic}, \ref{sec:binary}, and~\ref{sec:neutron_star}.

\newpage

\section{Theory and experiment frontier developments}
\label{sec:abstracts}

\subsection{Search for proton decay in the
Hyper-Kamiokande experiment}
\label{sec:HyperK}
\emph{Nicola Fulvio Calabria}

The discovery of proton decay would provide direct evidence of BSM physics.
The search for the two flagship decay modes $p \rightarrow e^{+}\pi^0$ and $p \rightarrow \nu K^+$ has been carried out in Super-Kamiokande, a 50-kton underground water-Cherenkov detector in operation since 1996 in the Kamioka mine, Japan. This has led to the establishment of the most stringent partial lifetime limits for both of them~\cite{Super-Kamiokande:2020wjk, Super-Kamiokande:2014otb}.

Hyper-Kamiokande is the next-generation underground water-Cherenkov detector currently under construction in the Tochibora mine, Japan, featuring a broad scientific program spanning from neutrino physics to the search for BNV. Hyper-Kamiokande is scheduled to start its operation in 2027, taking advantage of its huge fiducial mass ($186$ kton) and its new $20${\ttfamily"} photomultiplier tubes, which feature detection efficiency and timing resolution improved by a factor $2$ with respect to Super-Kamiokande, to achieve an unprecedented sensitivity to proton decay.

Hyper-Kamiokande is expected to probe proton partial lifetimes one order of magnitude higher than the best current limits, extending up to $10^{35}$ years~\cite{Hyper-Kamiokande:2018ofw}, thus playing a central role in exploring the future of particle physics.

\newpage
\subsection{Exploring baryon number violation at the Deep Underground Neutrino Experiment: a brief discussion of capabilities and possible future strategies}
\label{sec:DUNE}
\emph{Joshua L.~Barrow}

The Deep Underground Neutrino Experiment's (DUNE's) far detector complex of $4\times17\,$kt liquid argon time projection chambers (LArTPCs) will have a rich set of physics topics~\cite{DUNE:2020ypp}, including neutrino oscillation and BSM physics~\cite{DUNE:2020fgq}. Of great importance to the latter of these goals is BNV~\cite{Dev:2022jbf}, especially including nucleon decay~\cite{Alt:2020blf,Alt:2020rii,TylerStokesThesis}, $n\bar n$ transformations~\cite{Barrow:2021odz,Jwa:2022ocb,Jwa:2020mtz,MicroBooNE:2023dci,WheelerSIST}, and dinucleon decay, all of which hope to exploit DUNE’s particle calorimetry, low particle kinetic energy thresholds~\cite{Caratelli:2022llt,MicroBooNE:2017gxx}, and exceptional particle tracking capabilities with $\sim$mm-scale spatial resolution. All suffer from atmospheric neutrino backgrounds, which at times mimic these rare events' unique topologies. Here, we review recent results in this vein with a focus on proton decay (PDK) using the DUNE Far Detectors, as well as some of DUNE's principal experimental capabilities, and look forward to some possible upcoming analyses with new strategies discussed at the INT.

DUNE’s calorimetry and spatial resolution are key features to its future success in BNV searches, directly exploiting the topological characteristics~\cite{MicroBooNE:2023dci,Jwa:2022ocb,Jwa:2020mtz} of rare signals. Coupled with low charged particle kinetic energy thresholds (especially for charged hadrons~\cite{MicroBooNE:2017gxx}) and good particle identification via usage of $dE/dx$ and associated Bragg peaks---or lack thereof---near complete final states can \textit{in principle} be reconstructed. This said, much work continues within the collaboration on improvements to a plethora of reconstruction tools to better capture the full richness of complex final states.

Much attention has been paid within the DUNE collaboration to the predominately supersymmetric GUT-motivated mode of $p\rightarrow K^+ \bar{\nu}$, e.g., the ``golden channel'' of DUNE~\cite{DUNE:2020fgq,DUNE:2020ypp,Alt:2020blf,Alt:2020rii,TylerStokesThesis}, so-called due to LArTPC's very low kinetic energy thresholds for trackable charged hadrons compared to that of water-Cherenkov detectors~\cite{MicroBooNE:2017gxx}. Ideally, this channel leads to a topological signature of a highly-ionizing charged kaon before decaying a majority of the time into an antimuon, which subsequently decays into a Michel electron (positron). This triple track topology is hopefully reconstructed completely with visible and consecutive Bragg peaks. However, such a signal can in principle, among many other potential backgrounds, be confused with quasielastic-like charged current atmospheric neutrino scattering, wherein a highly-ionizing proton is produced in conjunction with a muon, which in principle could then decay into an electron if not captured. Such a background would similarly produce a triple track topology, although without consecutive connected Bragg peaks. Charged kaons of sufficient energies are indeed trackable within LArTPCs~\cite{Meddage:2019pbr,DUNE:2024qgl,ThorpeThesis,MicroBooNE:2022cls}, though it should be noted that the reconstructability of lower momentum kaons similar to those expected from intranuclear PDKs with momenta comparable to the Fermi momentum have not been widely studied within LArTPCs. Actively supporting and pursuing low-energy charged kaon beam studies could be considered to strengthen this understanding.

Of course, there are many other possible modes of BNV beyond ``golden channel'' PDK. Compared to its water-Cherenkov oriented competitors, DUNE is expected to shine not only in low-energy charged hadron tracking but more generally should excel in the reconstruction of high-multiplicity charged particle final states thanks to its spatial resolution and three-dimensional tracking. A classic example of such a search includes $n\bar n$ transformation, a specific form of the more general dinucleon decay, which is expected to rely on high-multiplicity pionic final states due to the antineutron's annihilation with an intranuclear nucleon partner~\cite{MicroBooNE:2023dci,Jwa:2020mtz,Jwa:2022ocb,Barrow:2021odz,LArIAT:2024lbx}. Exploiting the unique kinematics of such final states with novel reconstruction variables~\cite{WheelerSIST} is expected to be achievable within a LArTPC such as DUNE, hopefully enhancing the final sensitivity to such a search; being able to tag and reject charged current neutrino interactions as backgrounds is of course a key feature. Similarly, high-multiplicity leptonic final states could also be pursued with regard to proton or neutron decays, though no such sensitivity study has yet been performed.

Other possibilities for BNV searches are abundant, and search strategies could go beyond the mere enumeration of all relevant channels by the PDG~\cite{ParticleDataGroup:2024cfk}. Per discussions at the INT, consideration of (semi-)inclusive final states, such as $m N\rightarrow n e^\pm + X$, $m N\rightarrow n \mu^\pm + X$, $m N\rightarrow n \pi^\pm + X$, $\ldots$, $m N\rightarrow n \,\text{tracks} + X$, $\ldots$, $m N\rightarrow n \,\text{showers} + X$, etc., for $\{n,m\} = \{1,2,\ldots\}$, could also be pursued for continuous values of $X$ masses, wherein a continuous sensitivity limit could in principle be placed dependent upon identified particle kinematics. Given the (semi-)inclusive nature of such searches, background is expected to dominate, and so only lower-than-ideal limits would be expected to be placed, likely in merely a complementary manner. This said, setting individual limits on all such channels is of course possible. However, the scope of such (semi-)inclusive searches also points to the possibility of generic BNV searches as a form of ``anomaly'' detection within LArTPCs such as the DUNE far detectors, wherein powerful machine learning algorithms such as NuGraph~\cite{TylerStokesThesis,Aurisano:2024uvd} could be leveraged with broad binary classifiers alongside leading particle identification capabilities to select on BNV above atmospheric neutrino backgrounds.

\newpage
\subsection{Search for neutron--antineutron transition in large neutrino detectors}
\label{sec:nnbar}
\emph{Linyan Wan}

$n\bar n$ transition is a BNV process with $\Delta B=2$, providing a unique insight into the potential explanation of the baryon asymmetry in our Universe, especially in the context of post-sphaleron baryogenesis.
Studies have been conducted in various neutron-rich environment including free neutron sources~\cite{Baldo-Ceolin:1994hzw}, neutron stars~\cite{Goldman:2024yoh}, and bound neutrons in large neutrino detectors~\cite{Super-Kamiokande:2020bov, Super-Kamiokande:2011idx,Kamiokande:1986pyk,SNO:2017pha,Chung:2002fx,Frejus:1989csb,Irvine-Michigan-Brookhaven:1983nas}, neutron stars, and bound neutrons in large neutrino detectors, among which the large neutrino experiments~\cite{Super-Kamiokande:2020bov} obtained the most stringent constraints.

At Super-Kamiokande, we searched for $n\to\bar n$ oscillation with bound neutrons in the ${}^{16}$O nucleus of their water target, most recently with the full dataset from its first four run periods, representing an exposure of 0.37 megaton-years.
The search used a multivariate analysis trained on simulated $n\to\bar n$ events and atmospheric neutrino backgrounds featuring variables describing the kinematics, isotropy, particle multiplicity, and particle identification.
It resulted in 11 candidate events with an expected background of 9.3 events, in the absence of statistically significant excess, yielding a lower limit on $\bar n$ appearance lifetime in ${}^{16}$O nuclei of $3.6\times10^{32}$ years and on the $n\bar n$ oscillation time of $\tau_{n\bar n} > 4.7\times10^8$ s at 90\% C.L.~\cite{Super-Kamiokande:2020bov}.

Large future neutrino detectors such as Hyper-Kamiokande and DUNE are expected to push and extend the current limit and explore more into the theoretical predicted parameter space.
Hyper-Kamiokande, a next-generation water-Cherenkov detector, has a fiducial volume eight times that of Super-Kamiokande~\cite{Hyper-Kamiokande:2018ofw}, indicating a naively scaled up sensitivity of $\sqrt8\times3.6\times10^{32}$ years.
DUNE, a 40 kton detector utilizing liquid argon time projection technology, features a significantly lower detection threshold for hadrons compared to water-Cherenkov detectors.
This enhancement improves the efficiency for $n\to\bar{n}$ signals by a factor of 2 at the same background rate~\cite{DUNE:2020ypp}.
Ongoing efforts using machine learning reconstruction to study the topological features will provide additional insights.

\newpage
\subsection{Search for nucleon decay in JUNO}
\label{sec:JUNO}
\emph{Cailian Jiang}

The conservation of $B$ is an accidental symmetry in the SM of particle physics, and no fundamental symmetry guarantees the proton's stability. BNV is one of three basic ingredients to generate the cosmological matter--antimatter asymmetry from an initially symmetrical universe \cite{Sakharov:1967dj}. On the other hand, $B$ is necessarily violated, and the proton must decay in GUTs~\cite{Nath:2006ut}, which can unify the strong, weak, and electromagnetic interactions into a single underlying force.

JUNO is a 20 kton multipurpose underground liquid scintillator (LS) detector under construction in China, with a 650-meter rock overburden (1800 meter water equivalent) for shielding against cosmic rays \cite{JUNO:2015zny,JUNO:2021vlw, JUNO:2015sjr}. The LS detectors have distinct advantages in the search for some nucleon decay modes, such as $p\rightarrow \bar{\nu} K^+$ \cite{Undagoitia:2005uu,KamLAND:2015pvi,JUNO:2022qgr} and the neutron invisible decays \cite{Kamyshkov:2002wp,KamLAND:2005pen}.

\begin{figure}[!t]
	\centering
	\includegraphics[width=0.4\textwidth]{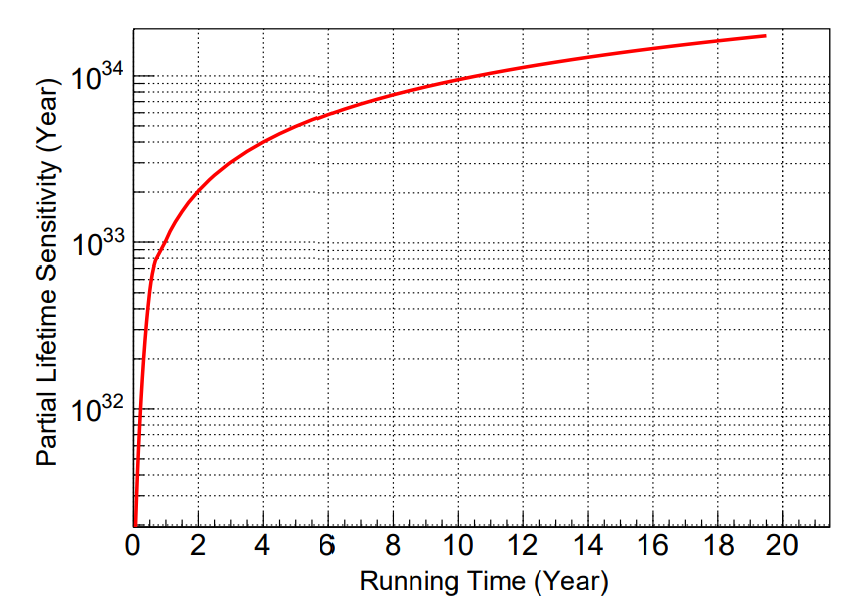}
 \includegraphics[width=0.4\textwidth]{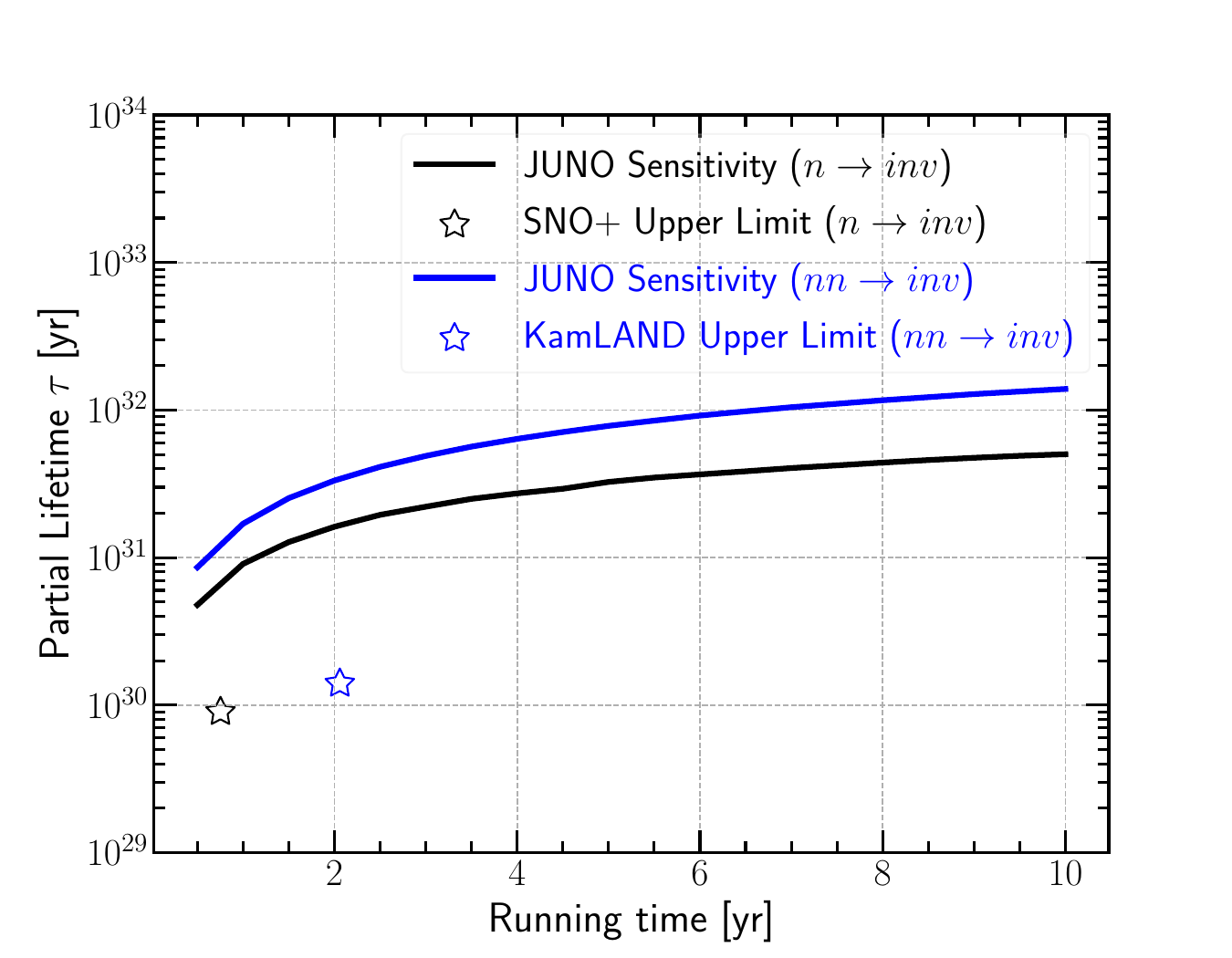}
    \caption{Left: the JUNO sensitivity for $p\rightarrow \bar{\nu} K^+$
as a function of running time at 90\% C.L. Right: JUNO sensitivities to $n \rightarrow \text{inv}$ and $n n \rightarrow \text{inv}$ as a function of the running time at 90\% C.L. SNO+ and KamLAND give the current best upper limits in the search of $n \rightarrow \text{inv}$ and $n n \rightarrow \text{inv}$ based on experimental data, respectively.}
	\label{fig:Tau_vs_running_time}
\end{figure}

The proton decay mode $p\rightarrow \bar{\nu} K^+$ is one of the two dominant decay modes predicted by a majority of GUTs \cite{Babu:2013jba}. In the JUNO LS detector, the decay will give rise to a three-fold coincidence feature in time, which is usually composed of a prompt signal by the energy deposit of $K^{+}$, a short-delayed signal ($\tau$ = 12.38 ns) by the energy deposit of decay daughters of $K^{+}$, and a long-delayed signal ($\tau$ = 2.2 $\mu$s) by the energy deposit of the final Michel electron. To discriminate the $p\rightarrow \bar{\nu} K^+$
signals from the enormous amount of backgrounds, we design a series of selection criteria based on the simulation data sample. The basic event selection about the visible energy is firstly applied. Then we employ all delayed signals selection of $p\rightarrow \bar{\nu} K^+$ and atmospheric neutrino events, including the Michel electron and neutron capture. Besides the common cuts on energy, position, and temporal features, additional criteria have to be explored since about 6.8\%
of the total atmospheric neutrino events would survive. The key part of the selections is based on the triple coincidence signature in hit time spectrum. We use the multi-pulse fitting method to reconstruct the time difference and energy of the $K^{+}$ and its decay daughters. After applying all criteria, the total efficiency for $p\rightarrow \bar{\nu} K^+$
is estimated to be 36.9\%, while the expected background level corresponds to 0.2 events in 10 years. Thus, the JUNO sensitivity on $p\rightarrow \bar{\nu} K^+$ at 90\% C.L.\ with 200 kton-years would be $\tau/\text{Br}(p\rightarrow \bar{\nu} K^+) > 9.6 \times 10^{33}$ years, which is higher than the current best limit of $5.9 \times 10^{33}$ yr from the Super-Kamiokande experiment \cite{Super-Kamiokande:2014otb}; see Fig.~\ref{fig:Tau_vs_running_time}(left).

We also explore the decay of bound neutrons into invisible particles (e.g., $n\rightarrow 3 \nu$ or $nn \rightarrow 2 \nu$) in the JUNO liquid scintillator detector, which do not produce an observable signal. The invisible decay includes two decay modes: $ n \rightarrow \text{inv} $ and $ nn \rightarrow \text{inv} $. The invisible decays of $s$-shell neutrons in $^{12}{\rm C}$ will leave a highly excited residual nucleus. Subsequently, some de-excitation modes of the excited residual nuclei can produce a time- and space-correlated triple coincidence signal in the JUNO detector. Based on a full Monte Carlo simulation informed with the latest available data,  we estimate all backgrounds, including inverse beta decay events of the reactor antineutrino $\bar{\nu}_e$, natural radioactivity, cosmogenic isotopes, and neutral current interactions of atmospheric neutrinos. Pulse shape discrimination and multivariate analysis techniques are employed to further suppress backgrounds. With two years of exposure, JUNO is expected to give an order of magnitude improvement compared to the current best limits. After 10 years of data taking, the JUNO expected sensitivities at 90\% C.L.\ are $\tau/\text{Br}(n \rightarrow \text{inv} ) > 5.0 \times 10^{31} \, {\rm yr}$ and $\tau/\text{Br}(nn \rightarrow \text{inv} ) > 1.4  \times 10^{32} \, {\rm yr}$; see Fig.~\ref{fig:Tau_vs_running_time}(right).

\newpage
\subsection{The HIBEAM and NNBAR projects at the European Spallation Source}
\label{sec:NNBAR}
\emph{John Womersley}

The European Spallation Source (ESS)~\cite{Garoby:2017vew} is now under construction in Sweden. Once operational, it will be the world's most intense neutron source. In addition to neutron scattering studies for materials and life sciences, the ESS has a dedicated mandate for fundamental physics research. A prioritization exercise in 2018 identified the absence of a dedicated beamline for particle physics as being a missing scientific capability of the highest importance, with the potential for exploring free $n\bar n$ oscillations at a level of sensitivity significantly beyond the current state of the art from ILL~\cite{Baldo-Ceolin:1994hzw}.

\begin{figure}[b!]
\centering
\includegraphics[width=0.8\textwidth]{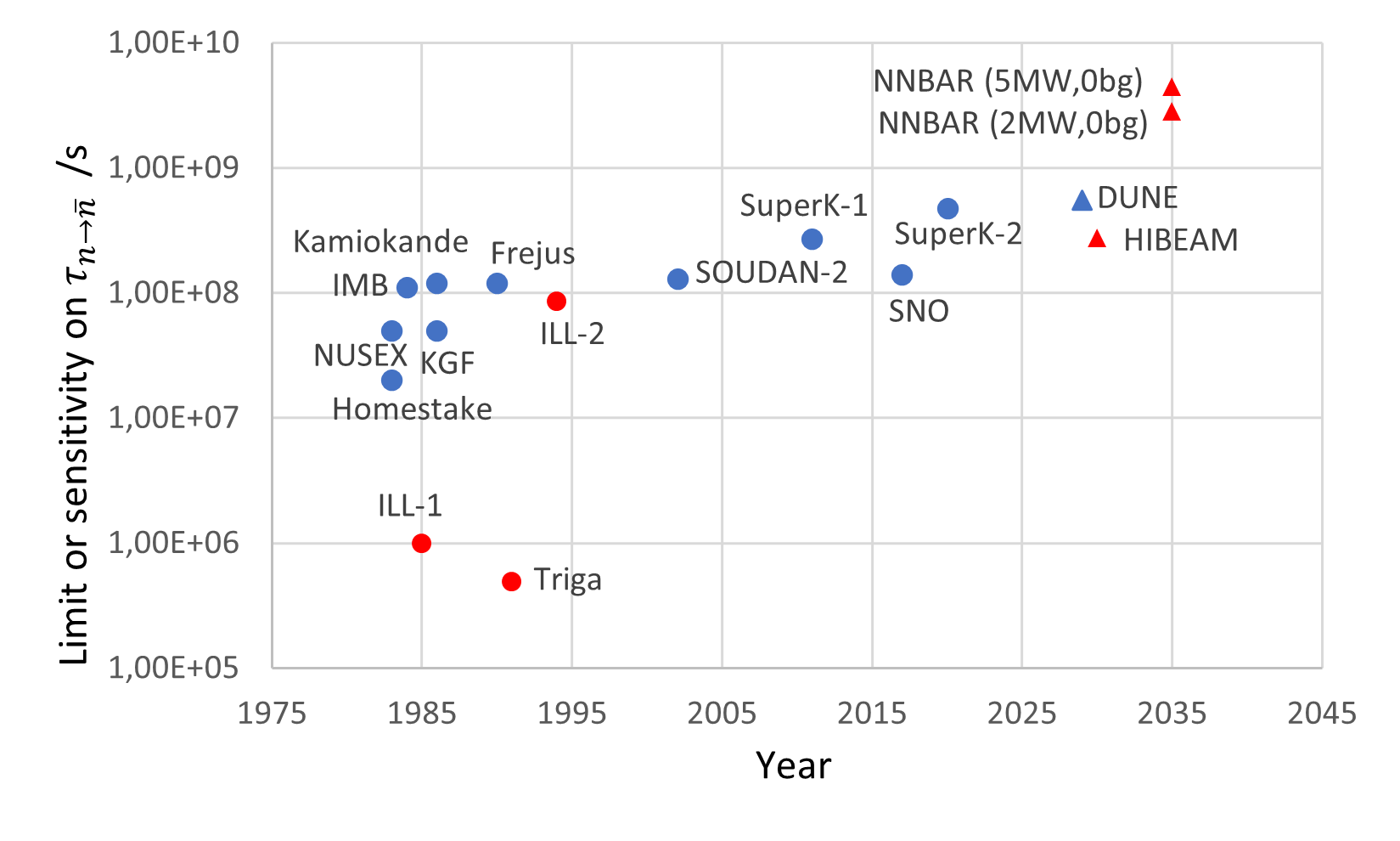}
\caption{Limits or sensitivity to the $n \rightarrow \bar n$ lifetime $\tau$ for a selection of published (circles) and planned (triangles) experiments. Blue symbols show model-dependent results from experiments using bound nucleons, and red symbols show results with free neutrons. The HIBEAM data point shows the expected sensitivity of HIBEAM with the WASA calorimeter at 5~MW.}
\label{sensitivity}
\end{figure}

A conceptual design for a new 200~m beamline and experiment NNBAR has been carried out~\cite{Backman:2022szk}. This would make use of a new large beam port and ideally a second moderator under the present ESS target wheel. Detailed MCNP and GEANT4 simulations of beamline and detector demonstrate that a sensitivity to observation of $n \rightarrow \bar n$ $\sim 1000 \times$ greater than ILL can be achieved. This clearly shows the potential for ESS to explore new regions of parameter space, but priorities and funding remain constrained in the short term. The chosen way forward is therefore to pursue a new, shorter particle physics beamline in the existing East experimental hall.  The High-Intensity Baryon Extraction and Measurement (HIBEAM) collaboration is now designing this beamline and related infrastructure.  The HIBEAM detector will make use of the existing WASA~\cite{CELSIUSWASA:2008vnq} crystal calorimeter to keep costs down.  Detailed modeling shows the potential to go a factor of 10 beyond ILL in discovery reach (a factor of $\sim 3$ in lifetime sensitivity) with 3--5 years of running at 5~MW, as shown in Fig.~\ref{sensitivity}.  HIBEAM can also explore other physics including searches for sterile neutrons and axion dark matter.

Currently, ESS council has approved 1.1 M\euro{} for the neutron extraction system. Preparatory support has been secured from the European Commission and the Swedish Research Council, and construction and testing of annihilation detector prototype components, validation and simulation, and system integration at the ESS test beam line are all underway.   The collaboration involves institutes from Sweden, USA, Israel, France, Italy, Brazil, Australia, with co-spokespersons G.~Brooijmans (Columbia) and D.~Milstead (Stockholm).  The project is ready to move forward when construction funding is secured
($\sim$15 M\euro{} is required for the minimum configuration).

\newpage
\subsection{A bottom-up approach to nucleon decay: RGEs, correlations, and connection to UV}
\label{sec:SMEFT}
\emph{Arnau Bas i Beneito}

Motivated by the upcoming series of experiments searching for proton decay~\cite{DUNE:2020ypp,Hyper-Kamiokande:2018ofw,JUNO:2015zny}, this work focuses on BNV from a bottom-up perspective. Specifically, we characterize nucleon decay in a model-independent framework based on SMEFT~\cite{Grzadkowski:2010es,Brivio:2017vri,Henning:2014wua}, highlighting the importance of searching for both $|\Delta (B - L)|=0, \;2$ nucleon decay modes. These modes are often indistinguishable from an experimental standpoint, yet their observation would point to different SMEFT operators, which originate from distinct UV models.

In particular, we analyze the lowest-dimensional BNV SMEFT operators of dimension six and seven~\cite{Weinberg:1979sa,Wilczek:1979hc,Abbott:1980zj,Claudson:1981gh,Lehman:2014jma}. We incorporate the RGE effects of such operators and perform a complete running procedure from the BNV scale, $\Lambda_{\slashed{B}}$, down to the proton mass scale, passing through the relevant EFTs~\cite{Alonso:2014zka,Liao:2019tep}. The nucleon decay rates are then computed using baryon chiral perturbation theory and the so-called indirect method~\cite{Aoki:2017puj,Yoo:2021gql},  with results that remain consistent with those obtained via the direct method \cite{Gargalionis:2024nij}. By leveraging the power of EFT, we derive model-independent limits on  $\Lambda_{\slashed{B}}$, consistently obtaining more stringent bounds when running effects are included.

Additionally, we present an analysis of the simultaneous presence of pairs of BNV operators, demonstrating how complementary searches in different decay channels can help break degeneracies. Particular attention is given to channels involving an $\eta$ meson in the final state. We also provide numerical expressions for the decay widths of various nucleon decay observables in terms of SMEFT Wilson coefficients at $\Lambda_{\slashed{B}}$,  assuming they are generated by integrating out heavy UV multiplets. The running effects are incorporated in all cases, and numerical coefficients for the decay rates are available online~\cite{kappa-matrices-zenodo,Beneito:2023xbk}, serving as a bridge between specific UV models and low-energy BNV observables.

Finally, we present a table of correlations among different BNV operators and the strength of their signals in various nucleon decay channels, normalized to the current experimental bounds on the proton lifetime~\cite{Super-Kamiokande:2020wjk,Super-Kamiokande:2014otb,Super-Kamiokande:2013rwg}. These results suggest that detecting signals in two or three channels could indicate specific operators and thus guide the search for viable UV completions, such as GUTs~\cite{Pati:1973uk,Georgi:1974sy,Fritzsch:1974nn}. The detection of the long-sought proton decay within the next decade would mark a groundbreaking milestone in particle physics. In such a scenario, identifying the operator responsible for the observed decay mode and investigating its predicted associated processes would be the primary course of action. Should that day arrive, the necessary ingredients for its analysis may be found here.

\newpage
\subsection{Two-loop anomalous dimensions of $\Delta B$ operators}
\label{sec:2loop}
\emph{Luca Naterop}

EFTs provide powerful tools in searches for BNV, as required to explain the baryon asymmetry of our Universe. Prominent theories, such as GUTs, differ by up to 15 orders of magnitude in energy from experimental approaches such as proton decay searches. A separation of scales by EFT methods is therefore crucial in order to improve perturbation theory through resummation of large logarithms, and for consistent combination of constraints across energy scales and observables.

Beyond the one-loop order, where the RGE of the relevant dimension-six $\Delta B$ operators is known \cite{Jenkins:2013zja,Jenkins:2013wua,Alonso:2013hga}, only partial results exist. The two-loop RGE available in the literature \cite{Nihei:1994tx} lacks proper scheme definition and includes only QCD corrections. But in a desired comprehensive EFT analysis of the proton decay, where bounds are evolved across vast energy scales, electroweak effects are expected to be of similar size due to approximate unification of forces.

Here, we explain an on-going effort towards a full determination two-loop $\Delta B$ RGE at dimension six, paving the way for comprehensive future EFT analyses. Even though the considered sector  does not lead to problems related to $\gamma_5$, the two-loop RGE is still scheme dependent. We discuss scheme dependencies that arise in $\Delta B$ sectors and we provide two well-defined schemes to be used: one in naive dimensional regularization and a second one in the 't Hooft--Veltman scheme for $\gamma_5$ \cite{tHooft:1972tcz}. Our schemes feature advantages such as: restoration of spurion chiral symmetry and proper definition and handling of evanescent operators \cite{Naterop:2023dek}. We also discuss a practical method for the calculation of the desired two-loop RGE. The method uses the well-known $R$ operation, which constructs all counterterms automatically, but must be adjusted to respected our schemes. The method has already proven useful in higher-loop EFT renormalization  calculations \cite{Naterop:2024ydo}. We find results~\cite{Naterop:2025lzc} that pass all consistency checks such as: gauge invariance, pole relations, and restoration of spurious chiral symmetry. The results are part of a broader community effort towards two-loop anomalous dimensions and, more generally, next-to-leading logarithmic EFT analyses \cite{Jenkins:2023rtg,Jenkins:2023bls,Born:2024mgz,Aebischer:2025hsx}.

\newpage
\subsection{How $A>1$ can help with $|\Delta B|=1,2$}
\label{sec:EFT_nuclear}
\emph{Ubirajara van Kolck}

Most experiments probe BNV through the decay of nuclei with the disappearance of one ($\Delta B=1$) or two ($\Delta B=2$) nucleons. The relation between a possible signal and in-vacuum processes, such as proton decay and $n\bar n$ oscillations, requires an understanding of the nuclear environment: effects of both $\Delta B =0$ and $\Delta B\ne 0$ internucleon interactions need to be assessed.

EFTs allow tracking $\Delta B\ne 0$ interactions from their manifestation in SMEFT as dimension-six ($|\Delta B|=1$) \cite{Weinberg:1979sa,Wilczek:1979hc,Abbott:1980zj} or nine ($|\Delta B|=2$) \cite{Kuo:1980ew,Rao:1982gt,Caswell:1982qs,Basecq:1983hi} operators at the electroweak scale $\sim 100$ GeV down to the nuclear scale $\sim 100$ MeV. At the lower scale, nuclear EFTs \cite{Hammer:2019poc} provide a systematic treatment of all internucleon interactions, with model independence ensured by
order-by-order renormalizability. The low-energy constants (LECs) can be determined from few-nucleon observables calculated with LQCD, and ``{\it ab initio}'' many-body methods can then be deployed to predict properties of larger nuclei \cite{Barnea:2013uqa}. In Chiral EFT,
the different chiral symmetry properties of $\Delta B\ne 0$ operators \cite{Claudson:1981gh,Buchoff:2015qwa} dictate the relative magnitudes of their contributions to various observables \cite{Oosterhof:2019dlo,Oosterhof:2021lvt}, enabling in principle the identification of a dominant operator, if it exists.

These virtues can be exemplified with the lifetime of the deuteron, which is dilute enough for pion exchange to be perturbative \cite{Kaplan:1998sz}:

\begin{itemize}
\item $\Delta B=1$ \cite{Oosterhof:2021lvt}: Once the energetic products of nucleon decay are integrated out, the LECs of $\Delta B=0$ nuclear interactions acquire imaginary parts. Up to and including next-to-leading order (NLO), the deuteron decay rate is given by the sum of the decay rates of the free proton and neutron. The first nuclear correction is expected to contribute at the few-percent level and comes with an undetermined LEC. These conclusions are in qualitative agreement with results from phenomenological potentials \cite{Dover:1981zj,Alvarez-Estrada:1982bkh}.

\item $\Delta B=2$ \cite{Oosterhof:2019dlo}: The relative importance of the $n\bar n$ transition and dinucleon decay depends on the $\Delta B=2$ operator at the quark level. For most operators, the ratio of the deuteron lifetime to the square of the $n\bar{n}$ oscillation time is fixed  at LO and agrees with results for short-range potentials \cite{Sandars:1980pr,Arafune:1981gw,Dover:1982wv,Kondratyuk:1996wq}. At NLO, there are calculable, small corrections and a so-far unknown short-range dinucleon LEC. Estimating this LEC with dimensional analysis, the result, which is analytical and has a quantified uncertainty, is smaller by a factor $\sim 2.5$ than estimates based on potential models \cite{Dover:1982wv,Haidenbauer:2019fyd}. For the remaining operator, there is no direct link to the $n\bar{n}$ oscillation time.

\end{itemize}
Details might change for heavier, denser nuclei, where the perturbative-pion expansion is expected to, at best, converge slowly. However, the main conclusions should survive~\cite{Oosterhof:2019dlo,Oosterhof:2021lvt}.

\newpage
\subsection{Indirect constraints on the third-generation baryon number violation}
\label{sec:third_gen}
\emph{Alexey A.~Petrov}

As baryon and lepton numbers are accidental symmetries of the SM, there is no reason to be conserved in general. The non-observation of BNV suggests that the scale of BNV interactions at zero temperature is comparable to the GUT scale. However, the pertinent measurements involve hadrons made of the first-generation quarks, such as protons and neutrons. One may, therefore, entertain the idea that new flavor physics breaks baryon number at a much lower scale, but only in the coupling to a third-generation quark, leading to observable BNV $b$-hadron decay rates. This motivation led to numerous searches for BNV decays of the $B$ mesons, $\Lambda_b$ baryons \cite{LHCb:2022wro,BaBar:2011yks}, and other heavy states \cite{Hou:2005iu,Crivellin:2023ter,Heeck:2024jei}.

We show \cite{Beneke:2024hox} that indirect constraints on the new physics scale $\Lambda_\text{BNV}$ from the existing bounds on the proton lifetime do not allow for this possibility. For this purpose, we consider the three dominant proton decay channels $p \to \ell^+ \nu_\ell \bar{\nu}$, $p \to \pi^+ \bar{\nu}$, and $p \to \pi^0 \ell^+$ mediated by a virtual bottom quark. To write a set of effective operators mediating this transition, we consider all BNV operators in SMEFT that contain at least one $b$-quark and match them into the Weak Effective Theory at the weak scale. We show that the operators with the left-handed $b$-quarks can yield the suppression of the order of $\sim |V_{ub}| \Lambda_{\rm GUT}$, so only the operators with the right-handed quarks should be considered.

Integrating out the $b$-quark, we are left with the effective Hamiltonian ${\cal H}_{\rm 6f} = \mathcal{H}_{p \to \ell^+ \nu_\ell \bar{\nu}} + \mathcal{H}_{p \to \pi^+ \bar{\nu}} +\mathcal{H}_{p \to \pi^0\ell^+}$ containing dim-9 operators that are local at the scale of 1 GeV,
\begin{align}
    \mathcal{H}_{p \to \ell^+ \nu_\ell \bar{\nu}} &= -2\sqrt{2}\,\frac{G_F C_\nu V_{ub}^*}{m_b \Lambda_{\rm BNV}^2} \,\mathcal{O}_{\nu,\text{sl}}  + \text{h.c.}\,,\nonumber\\
    \mathcal{H}_{p \to \pi^+ \bar{\nu}} &= -2\sqrt{2}\,\frac{G_F C_\nu V_{ub}^* V_{ud}}{m_b \Lambda_{\rm BNV}^2}\Bigl(C_1 \mathcal{O}_{\nu,1} + C_2 \mathcal{O}_{\nu,2} \Bigr) + \text{h.c.}\,,\notag\\
    \mathcal{H}_{p \to \pi^0\ell^+} &= -2\sqrt{2}\,\frac{G_F V_{ub}^*V_{ud}}{m_b \Lambda_{\rm BNV}^2}\sum_{X=L,R} C_X \Bigl(C_1 \mathcal{O}_{X,1} + C_2 \mathcal{O}_{X,2} \Bigr) + \text{h.c.}\,,
\end{align}
with the six-fermion operators
\begin{equation}
\label{eq:Onusl}
\mathcal{O}_{\nu, \text{sl}} = \varepsilon^{abc} [\widetilde{u}^a \gamma^\mu P_L u^b]\; [\widetilde{\nu} P_L d^c]\; [\bar{\nu}_\ell \gamma_\mu P_L \ell]\,,
\end{equation}
for $p \to \ell^+ \nu_\ell \bar{\nu}$, and
\begin{align}
\mathcal{O}_{\nu,1} &= \varepsilon^{abc}[\widetilde{u}^a \gamma^\mu P_L T^A_{bi}  u^i]\; [\widetilde{\nu}\,  P_L d^c]\; [\bar{u}^f \gamma_\mu P_L T^A_{fj} d^j]\,,\nonumber\\
\mathcal{O}_{\nu,2} &= \varepsilon^{abc} [\widetilde{u}^a \gamma^\mu P_L u^b]\; [\widetilde{\nu}\,  P_L d^c]\; [\bar{u}^f \gamma_\mu P_L d^f]\,,\nonumber\\
\mathcal{O}_{X,1} &= \varepsilon^{abc}[\widetilde{u}^a \gamma^\mu P_L T^A_{bi}  u^i]\; [\widetilde{\ell}\,  P_X u^c]\; [\bar{u}^f \gamma_\mu P_L T^A_{fj} d^j]\,,\nonumber\\
\mathcal{O}_{X,2} &= \varepsilon^{abc} [\widetilde{u}^a \gamma^\mu P_L u^b]\; [\widetilde{\ell}\,  P_X u^c]\; [\bar{u}^f \gamma_\mu P_L d^f]\,,
\end{align}
for $p \to \pi^+ \bar{\nu}$ and $p \to \pi^0 \ell^+$. Note that the tree-level matching coefficient is $1/m_b$ for all operators,  and the subscripts refer to the BNV and weak currents, respectively, and $X=L, R$.  Computing the matrix elements of dim-9 operators and comparing the results to the experimental constraints from $p \to \ell^+ \nu_\ell \bar{\nu}$, $p \to \pi^+ \bar{\nu}$, and $p \to \pi^0 \ell^+$, we conclude that for all transitions $\Lambda_\text{BNV}/\sqrt{|C_i|} \gtrsim \text{a few} \times 10^9$ GeV  \cite{Beneke:2024hox}, which implies that the obtained lower limits on the scale of highly generation-dependent BNV favoring the third family {\it exclude} the possibility of observing such BNV directly in $b$-hadron decays in any presently conceivable experiment.

\newpage
\subsection{Proton decay matrix element on the lattice}
\label{sec:lattice_proton_decay}
\emph{Jun-Sik Yoo}

Proton decay is a major prediction of GUTs \cite{Pati:1973uk,Georgi:1974sy,Fritzsch:1974nn,Weinberg:1979sa,Wilczek:1979hc,Abbott:1980zj,Espinosa:1989qn,Ambjorn:1990pu}, and its observation would indicate BNV, a key requirement for baryogenesis \cite{Sakharov:1967dj}. However, despite extensive searches spanning several decades, no definitive proton decay signal has been observed, leading to stringent constraints on its rate and ruling out some of the simplest GUT models. The decay rate depends not only on BNV interactions, but also on the hadronic matrix elements that encode the transition amplitudes between the proton and the final-state mesons and antileptons. These matrix elements, which arise from three-quark operators, are inherently nonperturbative and require first-principles calculations to be determined accurately.

We present results from a nonperturbative LQCD calculation of the proton decay matrix elements for the most commonly studied two-body decay channels into a meson and an antilepton \cite{Yoo:2021gql}. Our simulations are conducted with physical light and strange quark masses on lattices with spacings $a \approx 0.14$ fm and $0.20$ fm, allowing us to systematically control uncertainties related to discretization and quark mass dependence. In addition to computing the direct three-quark matrix elements, we also evaluate the proton-to-vacuum matrix elements, $\alpha$ and $\beta$, which serve as important inputs for phenomenological models of proton decay. We employ nonperturbative renormalization to match our results to continuum schemes, ensuring robust predictions. Furthermore, we perform a careful excited-state analysis to mitigate contamination from higher-energy states, improving the reliability of our results.

Our findings largely agree with previous LQCD calculations that used heavier quark masses \cite{Hara:1986hk,JLQCD:1999dld,CP-PACS:2004wqk,Aoki:2008ku,QCDSF:2008qtn,Aoki:2013yxa,Aoki:2017puj}, reinforcing the conclusion that the quark mass dependence of hadron structure does not significantly alter the exclusion of certain simple GUT models \cite{Hisano:1992jj,Murayama:2001ur}. Moreover, our study lays the groundwork for extending these calculations to other possible proton decay channels, further broadening the scope of LQCD contributions to constraining GUT scenarios. Future work will explore additional decay modes and extend the precision of these matrix element computations by incorporating finer lattice spacings and larger volumes.

\newpage
\subsection{Towards baryon number violation in nuclei from lattice QCD}
\label{sec:lattice_nnbar}
\emph{Michael L.~Wagman}

$n\bar n$ oscillations, which violate $B$ and $B-L$, are predicted as low-energy signatures of BSM post-sphaleron baryogenesis scenarios~\cite{Babu:2006xc,Babu:2013yca}.  Experimental bounds on the $n\bar{n}$ transition time, $\tau_{n\bar{n}}$, have been obtained from searches for free neutron transitions~\cite{Baldo-Ceolin:1994hzw} as well as intranuclear $n\bar{n}$~\cite{SNO:2017pha,Super-Kamiokande:2020bov}. New discoveries or limits will be set for free neutrons at the ESS~\cite{Santoro:2023izd} and for intranuclear $n\bar{n}$ by Hyper-Kamiokande and DUNE~\cite{Barrow:2019viz}.

Relating $\tau_{n\bar{n}}$ to BSM parameters is accomplished in SMEFT by introducing a complete basis of six-quark operators, $Q_I$~\cite{Rao:1982gt}.
Predictions for particular BSM theories are encoded in the Wilson coefficients $C_I$ of the Hamiltonian $\mathcal{H}_{n\bar{n}} = \sum_I C_I Q_I$.
Its matrix element gives the $n\bar{n}$ transition rate $\tau_{n\bar{n}}^{-1} =  \left| \left< \bar{n} | \mathcal{H}_{n\bar{n}} | n \right> \right|$.
LQCD has been used to compute the $\left< \bar{n} | Q_I | n \right>$ in a convenient basis~\cite{Buchoff:2015qwa} where the $Q_I$ renormalize multiplicatively and there are only four non-zero matrix elements~\cite{Rinaldi:2018osy,Rinaldi:2019thf}.
For example, in the left-right symmetric model of Ref.~\cite{Mohapatra:1980qe}, the ILL bound and these LQCD results provide a bound of $350(22)$ TeV on the scale of $B$ violation.

Further complications arise in relating $\tau_{n\bar{n}}$ to intranuclear $\Delta B = 2$ rates.
Chiral EFT has been used to relate the deuteron lifetime $\Gamma_d^{-1}$ constrained by SNO to $\tau_{n\bar{n}}$ using Kaplan--Savage--Wise (KSW) power counting in Ref.~\cite{Oosterhof:2019dlo} and using Weinberg power counting in Ref.~\cite{Haidenbauer:2019fyd}.
The results differ by a factor of two, indicating that higher-order effects (in at least one power counting) are significant.
To constrain these higher-order Chiral-EFT effects, LQCD calculations of $n\bar{n}$ in multi-nucleon states are needed.

The simplest multi-nucleon matrix element of $Q_I$ for LQCD is $\left<0| Q_I | nn \right>$, with $nn$ being the two-neutron ground state. The corresponding quark-line diagram is simply the crossed version of the diagram describing $n\bar{n}$.
Further, this matrix element can be identified as an ``overlap factor'' in a two-point correlation function involving $Q_I$ and an operator for the $nn$ ground state.
Conveniently, such two-point functions were recently computed
in variational studies of the $nn$ spectrum
by the NPLQCD Collaboration in Ref.~\cite{Detmold:2024iwz}.
The $Q_I$ operator does not appear to couple strongly to the $nn$ ground state and has significantly larger overlap with the states coupling to plane-wave product operators $n(\mathbf{k}) n(-\mathbf{k})$ with non-zero $\mathbf{k}$.
This suggests that intranuclear $n\bar{n}$ rates could have significant dependence of the neutron momentum distributions of nuclei.

Current LQCD results only constrain the bare $\left<0| Q_I | nn \right>$, and quantitative results must wait for determinations of the corresponding renormalization factors.
Further challenges associated with energy-eigenstate identification for multi-nucleon systems can be mitigated using a new analysis framework based on the Lanczos algorithm that can provide rigorous two-sided bounds on the systematic errors associated with excited-state contamination~\cite{Wagman:2024rid,Hackett:2024xnx,Hackett:2024nbe}.
Complete LQCD determinations of $\left<0| Q_I | nn \right>$, for two-neutron ground and excited states, could then be matched to Chiral EFT to constrain the relation between $\Gamma_d^{-1}$ and $\tau_{n\bar{n}}$.
After matching LQCD and  Chiral EFT in the two-nucleon sector, the same relations would inform Chiral-EFT predictions relating intranuclear $n\bar{n}$ rates to $\tau_{n\bar{n}}$ in larger nuclei.

\newpage
\subsection{Lattice QCD computations of proton decay matrix elements}
\label{sec:lattice_proton_decay_2}
\emph{Yasumichi Aoki}

Proton decay matrix elements for the final states of one lepton and a pseudoscalar meson are computed with LQCD utilizing physical-point ensembles of the PACS collaboration using nonperturbatively ${\mathcal O}(a)$ improved Wilson fermions. The form factors necessary to compute all of these processes with all possible three-quark operators are studied. The latest but preliminary results are available in Ref.~\cite{Tsuji:2025xvq}. This computation uses finer lattices than Ref.~\cite{Yoo:2021gql}, at the cost of violating lattice chiral symmetry. With  different characteristics, two computations would provide a good crosscheck.
In the following we sketch important ingredients for the lattice computation.

Necessary steps for computing the matrix element in a certain renormalization scheme and scale are 1) computation of the matrix elements from three-point functions and 2) renormalization and matching of the BNV three-quark operators.
There is a well-established way for 1), first done by JLQCD \cite{JLQCD:1999dld}.
The difference from that to the current state-of-the-art computation is the use of dynamical fermion ensembles generated for the physical average up- and down-quark mass and physical strange-quark mass.
2) is done using the Rome--Southampton, RI/MOM~\cite{Martinelli:1994ty} (regularization independent, momentum subtraction) nonperturbative renormalization scheme constructed for proton-decay operators~\cite{Aoki:2006ib}.
The RI/MOM scheme acts as an intermediate renormalization to bridge the lattice results to more convenient renormalization schemes like $\overline{\text{MS}}$, and is designed to absorb the lattice discretization details. The rest, matching to, e.g., $\overline{\text{MS}}$, can be done in continuum perturbation theory.

There are several different RI/MOM schemes  used so far. For the quark wavefunction renormalization we typically utilize the vertex functions of bilinear operators with vanishing anomalous dimension, such as vector and axial vector operators. To avoid contamination of IR physics effect, RI/SMOM schemes \cite{Aoki:2007xm,Sturm:2009kb,Almeida:2010ns}  making use of non-exceptional momenta are applied for the bilinear vertex functions.

For the three-quark vertex, in the original RI/MOM scheme \cite{Aoki:2006ib},
all three momenta flowing in from three external quark legs are set equal ($p$), which makes the momentum flow from the three-quark operator $3p$. A new scheme discussed in Ref.~\cite{Gracey:2012gx}, which is used in Refs.~\cite{Yoo:2021gql,Tsuji:2025xvq}, sets the momentum flow from the three-quark operator to zero. In these schemes there is an imbalance in the momenta involved: $p$ vs.\ $3p$ or $p$ vs.\ zero. In Ref.~\cite{Yoo:2021gql} the old and new schemes are called MOM$_{3q}$ and Symm$_{3q}$.
There is a better choice of the momentum configuration following the discussion in Ref.~\cite{Aoki:2007xm}, where the sizes of all four momenta are set equal. The matching computation for such a scheme is reported in Ref.~\cite{Kniehl:2022ido}, but has not been used in LQCD computations yet.

One combination of the schemes of the bilinear vertex and the three-quark vertex defines one scheme for the three-quark operator. As the perturbative matching factor to a given order of $\alpha_s$ for one combination differs from another, multiple schemes constructed in such a way can be used for the estimate of the truncation error of perturbation theory.

\newpage
\subsection{Opening up baryon-number-violating operators}
\label{sec:opening}
\emph{Julian Heeck}

As shown by Weinberg~\cite{Weinberg:1979sa}, BNV first arises at mass dimension $d=6$ in SMEFT and leads to clean $\Delta B = \Delta L=1$ two-body nucleon decays, even if the operators contain second- or third-generation fermions~\cite{Hou:2005iu,Beneke:2024hox,Heeck:2024jei}. Most, but by far not all, of these decays are under experimental scrutiny at Super-Kamiokande and will be further improved with the upcoming detectors Hyper-Kamiokande and DUNE~\cite{FileviezPerez:2022ypk}.

Alas, as also emphasized by Weinberg~\cite{Weinberg:1980bf}, the naively-dominant $d=6$ operators could be suppressed or even forbidden if $U(1)_B\times U(1)_L$ is broken in non-trivial directions~\cite{Heeck:2019kgr}. For example, if $B+L$ remained conserved, the $d=7$ $\Delta B=-\Delta L=1$ operators would dominate, leading to \textit{different} two-body nucleon decays; if $U(1)_L$ is conserved, the $d=9$ $\Delta B = 2$ operators dominate, leading to $n\bar{n}$ transitions, etc.
The unparalleled sensitivity of nucleon-decay experiments allows them to see effects from $d\gg 6$ operators, which could indeed dominate due to their special $U(1)_B\times U(1)_L$ numbers. Since $d\gg 6$ $\Delta B$ operators generically lead to multi-body final states, it is more challenging to cover them experimentally. A few dedicated searches~\cite{Super-Kamiokande:2018apg,Super-Kamiokande:2020tor} for particularly clean $d\geq 10$ operators~\cite{Hambye:2017qix} have been performed, but exhaustive coverage along these lines appears unfeasible. \textit{Inclusive} searches could provide broad coverage with comparatively minimal effort~\cite{Heeck:2019kgr}.

Imposing $U(1)_B\times U(1)_L$ subgroups or flavor symmetries  is not the only way to push $\Delta B$ to $d>6$: as yet again shown by Weinberg~\cite{Weinberg:1980bf}, new particles sometimes only generate a select few $d> 6$ $\Delta B$ operators even without imposing \textit{any} symmetries. In an effort to systematically explore such \textit{accidentally protected operators}, we have constructed tree-level UV completions of all non-derivative $d\leq 14$ $\Delta B$ operators~\cite{Heeck2025}. We can then find all UV completions of an operator $\mathcal{O}$ at a given  $d$ that do not generate $\Delta B$ at lower $d$. If $\mathcal{O}$ carries $(\Delta B,\Delta L)$ that do not appear at lower $d$, this is just a UV completion with an \textit{accidental} $U(1)_B\times U(1)_L$ symmetry and the same phenomenology as the symmetry-protected operators. However, if  $\mathcal{O}$ carries the same $(\Delta B,\Delta L)$ as lower-$d$ operators, those will be induced through loops, automatically providing all genuine/finite loop $\mathcal{O}_{\Delta B}$ realizations. Importantly, it is not always clear that the loop-induced operators dominate the phenomenology, despite their lower $d$, because they can come with additional suppression factors, e.g., Yukawa couplings. A careful study of these operators is thus  necessary.
The surprising outcome of our study up to $d=12$ so far is that almost \textit{all} $\Delta B$ operators can be realized in this way, just by picking the right UV completion! Theoretically, almost \textit{any} $\Delta B$ operator at \textit{any} mass dimension could serve as the starting point, vastly enlarging the landscape of interesting $\Delta B$ operators.

Despite half a century of work, we have but scratched the surface of BNV. With the newly established vast landscape of potentially-dominant $\Delta B$ operators and models, it is now up to the theorists to calculate the signatures (this includes operators with more than three quarks and eventually also derivatives) and experimentalists to efficiently search for the new channels.

\newpage
\subsection{Scalar leptoquarks in loop-induced proton decays}
\label{sec:leptoquark}
\emph{Svjetlana Fajfer}

Baryon number is conserved in the SM due to an accidental symmetry, while in many BSM approaches  baryon number is violated.
Wigner already, in 1949, proposed to test baryon number conservation \cite{Goldhaber:1980dn}. Since then, several experiments have bounded the proton lifetime, with the most recent bounds coming from the Super-Kamiokande experiment, which found that the proton is stable up to $10^{34}$ years \cite{ParticleDataGroup:2024cfk}.
The lifetime of the decay $p \to \pi^0 e^+$ is the most constrained among all decay modes \cite{FileviezPerez:2022ypk}.  This decay mode was investigated from the eighties  \cite{Weinberg:1979sa,Wilczek:1979hc,Weinberg:1980bf} till today (see, e.g., Refs.~\cite{Nath:2006ut,Heeck:2019kgr}). The authors of Refs.~\cite{Gargalionis:2024nij,Beneito:2023xbk,Heeck:2019kgr} relied on
the effective Lagrangian approach instead of using a specific model. Within this approach,  operators can have the dimension six or higher  \cite{Gargalionis:2024nij,Dorsner:2022twk,Heeck:2019kgr,Beneke:2024hox,Dorsner:2012nq}.
Some operators result from the loop diagrams \cite{Gargalionis:2024nij,Dorsner:2022twk,Dorsner:2012nq}.
In particular,  heavier quarks or leptons mediate the BNV process at the loop level \cite{Beneke:2024hox}.

In addition to the box diagrams  \cite{Dorsner:2012nq}, we have noticed that the triple-leptoquark interactions might generate proton instability for two different decay topologies under the assumption that scalar leptoquarks of interest couple solely to the quark--lepton pairs \cite{Dorsner:2022twk}.
The first topology has the tree-level structure and yields three-body proton decays at leading order \cite{Dorsner:2022twk}. The other topology, one-loop, generates two-body proton decay processes instead. The tree-level topology has been previously analyzed in the literature (see, e.g., Refs.~\cite{Kovalenko:2002eh,Klapdor-Kleingrothaus:2002rvk,Arnold:2012sd,Murgui:2021bdy}),  while the one-loop level one has not been studied in the former studies.
Interestingly, we find out that the one-loop level topology produces more stringent bounds on the scalar leptoquark masses than the tree-level topology branching ratio bounds  \cite{Dorsner:2022twk}. Using the latest experimental input \cite{ParticleDataGroup:2024cfk}, we determined a lower limit on the mass scale associated with the leading-order proton decay signatures for both topologies within one of several possible scenarios.
We found the limit on this scale for the one-loop level process $p \to \pi^0 e^+$ to be $ \Lambda  \geq 1.8 \times 10^4$ TeV when the charged lepton in the loop is an electron. From the tree-level topology process $p \to e^+ e^+ e^-$, we determined that  $ \Lambda  \geq 1.2 \times 10^2$ TeV.  We assume that the scale $ \Lambda$  is equal to the masses of all those leptoquarks that participate in a given BNV process and set all of the dimensionless couplings to one. Also, we relied on the assumption that leptoquarks couple solely to the first-generation SM fermions.
We also specified the most prominent proton decay signatures due to the presence of all non-trivial cubic and quartic contractions involving three scalar leptoquark multiplets, where in the latter case, one of the scalar multiplets is the SM Higgs doublet.

 The decay widths for the nucleon to an antilepton and a photon are not so well bounded experimentally as nucleon decays to a pion and lepton. Using an effective Lagrangian approach, we related the decay widths of $N\to \ell \gamma $ to the decay widths of $N\to \ell \pi$.
Our model-independent result indicated  a factor $10^{3}$ suppression of the decay widths $\Gamma (p   \to \ell^{+} \gamma)$ and $\Gamma (n \to \bar \nu  \gamma)$ compared to the decay widths $\Gamma (p  \to \ell^{+}  \pi^{0})$ and $\Gamma (n \to \bar \nu \pi^{0})$, respectively \cite{Fajfer:2023gfi}.

\newpage
\subsection{Baryon violation in cosmic rays: UHECR and cosmic antinuclei}
\label{sec:cosmic}
\emph{Zurab Berezhiani}

Dark matter in the Universe may exist in the form of mirror (M) matter
represented by the parallel hidden sector of  particles
which is an exact replica of ordinary (O) particle sector.
Then all O particles: electron $e$, proton $p$, neutron $n$, neutrinos $\nu$, etc.\ must
have their M twins $e', p', n', \nu'$  etc., which are sterile to the SM interactions, but have their own SM$'$  gauge interactions (for a review, see Ref.~\cite{Berezhiani:2003xm}).
Cosmological bounds require M sector to be colder than O sector, $x=T'/T \ll 1$,
which can be realized by asymmetric reheating  \cite{Berezhiani:1995am}
and by demanding the out-of-equilibrium
condition on cross-interactions between the O and M particle species
at post-inflation epoch  \cite{Berezhiani:2000gw}.
Namely, M baryons pass the cosmic-microwave-background and large-scale-structure tests
if $x< 0.2$ or so \cite{Berezhiani:2003wj}.
In M sector, being colder and helium-dominated, stars are formed earlier
and evolve faster \cite{Berezhiani:2005vv}, and their collapse can produce
heavy black holes  which  may dominate galactic halos along with M stars in solar mass range.

A specific feature of this scenario is that any neutral O particle, elementary (e.g., neutrinos)
or composite (as, e.g., the neutron) can have a mixing with its mass degenerate M twin.
Such mixings can be induced by $L$- or $B$-violating cross-interactions
between O and M leptons and quarks whose interactions, remarkably,
can  generate  baryon asymmetries in both sectors.
In particular, the same mechanism which induces  $\nu$--$\nu'$ mixings and makes
M neutrinos natural candidates for sterile neutrinos \cite{Akhmedov:1992hh,Berezhiani:1995yi},
also suggests the co-leptogenesis scenario \cite{Bento:2001rc}
which naturally explains the cosmological concordance  between the ordinary and dark
matter fractions in the Universe, $\Omega_{\text{B}'}/\Omega_{\text{B}} \simeq 5$
\cite{Berezhiani:2008zza},  and also predicts that the baryon asymmetry in M sector must have 
the sign opposite to that of O sector \cite{Berezhiani:2018zvs}.

As for the mass mixing term  $\epsilon \bar{n}n' + \text{h.c.}$ between the neutron and M neutron,
the present experimental data allow a maximal $n$--$n'$ oscillation in vacuum with a characteristic
time $\tau_\epsilon= \epsilon^{-1}$  much shorter than the neutron decay time  $\tau_n \approx 880$~s
\cite{Berezhiani:2005hv,Berezhiani:2009ldq,Berezhiani:2020vbe}.
The effects of $n$--$n'$ oscillation strongly depend on background conditions,
and it  could explain the neutron lifetime anomaly  \cite{Berezhiani:2018eds}.
It may also have interesting astrophysical consequences,
in particular for the ultra high energy cosmic rays (UHECR) \cite{Berezhiani:2006je,Berezhiani:2011da}
and for the neutron stars (NS) \cite{Berezhiani:2020zck}.
These observations triggered intense search of $n$--$n'$ transitions at the
cold and ultra-cold neutron facilities. Results of several dedicated experiments
are summarized in Refs.~\cite{Berezhiani:2017jkn,nEDM:2020ekj}.
They set lower bounds on $\tau_\epsilon$, typically  $1\text{--}100$ seconds
depending on experimental conditions.
Situation remains controversial:  some of experiments
show from $3\sigma$ to
5$\sigma$ deviations from null-hypothesis.

Here I discuss recent astrophysical hints which would be hard to explain
without invoking $n$--$n'$ mixing. These are the following:
(A)  two years ago it was reported that the AMS-2 experiment had observed
10 candidate events for anti-helium nuclei in cosmic rays~\cite{Ting:2023}. Discovery of a single anti-nucleus
with $A\geq 4$, if confirmed definitely, would challenge all known physics
and cosmology, and its impact
can be comparable to that of the positron discovery in 1932.
(B) paradoxical situation in the UHECR: cosmic ray fluxes detected by two giant experiments,
Pierre Auger Observatory (PAO) mostly exposed to South Hemisphere
and Telescope Array (TA) exposed to North Hemisphere,
are perfectly compatible at energies below the GZK cutoff  $E_\text{GZK}
\simeq 50$~EeV but become discrepant at higher energies, most notably at
$E > 100$ EeV.
In addition, since the propagation length of such energetic UHECR is typically
$\ell \sim$ 10 Mpc, they should come from  potential
sources closer than 30 Mpc or so. However, their arrival directions have no  correlations with
the nearby sources, and  it seems that they rather come from the most empty regions of the sky.
In particular, the extreme energy event ($E=244$~EeV) recently detected by
TA collaboration \cite{TelescopeArray:2023sbd} points  towards the center
of the Local Void.

These phenomena can be explained   as follows:

(A)  In fact, $n\to n'$ conversion can gradually transform a NS into a mixed star
having the core of M matter \cite{Berezhiani:2020zck}. The transformation time
can be estimated as $t_\epsilon \sim (\tau_\epsilon/1\,\text{s})^2 \times 10^{15}$~yr,
while constraints from the NS internal heating imply  $t_\epsilon > 10^{15}$~yr or so.
Thus, a typical old NS of mass $M\simeq 1.5\,M_\odot$ of the age $t\sim 10$ Gyr
would produce  $N_{B'} = (t/t_\epsilon) \times 2\times 10^{57}$ mirror baryons  in its interior.
If $B' <0$, the mirror NS (NS$'$) should consist of $\bar{n}'$,
and the reverse process $\bar{n}' \to \bar{n}$
would produce the same amount $N_{\bar B}$ of ordinary antibaryons in their interiors.
The latter should form, via nucleosynthesis processes, a core of anti-nuclei which can be
released after the gravitational mergers of the mirror NS \cite{Berezhiani:2021src}.
The cosmic anti-baryons injection rate, 
$2 N_{\bar B}\, R''$,
by taking  $N_{\bar B}\sim 10^{53}$ and adopting the  LIGO rate
$R' \sim 10^3\,\text{Gpc}^{-3}\,\text{yr}^{-1}$
for the NS$'$--NS$'$ mergers, 
nicely fits the anti-helium flux observed by AMS-2.

(B) As far as M sector is colder, $x=T'/T < 0.2$ or so, the UHECR
from M sources practically do not suffer the GZK cutoff and their flux
can be 2--3 orders of magnitude larger than that of ordinary ones,
since their propagation length scales as $\ell' \simeq \ell/x^3$ and it can be as large as
1 Gpc or so \cite{Berezhiani:2006je}.
Nevertheless, mirror UHECR with energies $E > x^{-1} E_\text{GZK}$, 
in their collisions with mirror relic photons, can produce the M neutrons $n'$  which, provided that $\tau_\epsilon \ll \tau_n$, can promptly oscillate into ordinary neutrons $n$ and then produce protons via $\beta$ decay $n\to pe \bar{\nu}$. 
The energetic protons produced in this way  
within the ordinary GZK radius 
(i.e., in the nearby sky at distances $d < 30$ Mpc or so), 
can safely reach us \cite{Berezhiani:2011da}.
Interestingly, the condition for $n'$--$n$ oscillation to not be effective
reads as $(\tau_\epsilon/1\,\text{s}) (B/1~\text{fG}) < 1$, and it 
can be satisfied only in cosmic voids where magnetic  fields can be as small as $B < 1$~fG.
This would naturally explain why the most of extreme energy events come from the voids.
In addition, since the Northern Sky is more ``voidy,'' and 
namely the Local Void is mostly exposed to the TA experiment, this can also explain the difference between the PAO and TA fluxes at $E> 100$ EeV.

 Summarizing, the situation looks very optimistic for the ongoing and planned searches at different neutron
 facilities at the PSI, ILL, ORNL, and ESS  \cite{Ayres:2021zbh,Addazi:2020nlz,Abele:2022iml}.
 The NS bounds on $n$--$n'$ conversion imply $\tau_\epsilon > 1$~s or so \cite{Berezhiani:2020zck},
 and in this case production of antimatter via $\bar{n}'\to \bar{n}$ in the cores of mirror NS
 can explain  the paradox (A) on the cosmic anti-helium flux.
 On the other hand, solution of the paradox (B) quests for $\tau_\epsilon < 100$~s or so.
 This gives a hope that under the systematic search
 these experiments may indeed discover $n$--$n'$ oscillation
 in the range  $\tau_\epsilon = (1\text{--}100)$s.

\newpage
\subsection{Some recent results on $n\bar n$ transitions}
\label{sec:nnbar_theory}
\emph{Robert Shrock }

The violation of baryon number is one of the necessary
conditions for baryogenesis.  In addition to proton decay, an
interesting possibility for BNV is $n\bar n$ transitions
\cite{Mohapatra:1980qe}. Although the SM conserves
$B$ perturbatively, BNV is present in many BSM
theories. Here, we first review the basic formalism describing
$n\bar n$ transitions in quasi-free neutron propagation experiments
and in matter, and then discuss a model in which proton decay can be
suppressed well below current experimental limits while $n\bar n$
transitions and associated $\Delta B=-2$ dineutron decays can occur
near to current limits
\cite{Nussinov:2001rb,Girmohanta:2019fsx,Girmohanta:2020qfd}. Thus, in
this model, $n\bar n$ transitions and associated $\Delta B = -2$
dineutron decays are the main manifestation of BNV, rather than proton decay. This model features a certain number of extra
compactified spatial dimensions, with left- and right-handed chiral
components of SM fermion fields having wave functions that are
strongly localized at various different points in the extra
dimensions. It has the appeal that it can naturally explain the
generational hierarchy in quark and charged lepton masses
\cite{Arkani-Hamed:1999ylh}, and can also explain the masses and
mixings of neutrinos \cite{Girmohanta:2020llh}. The effective 4D
Lagrangian and coefficients of six-quark operators responsible for the
$n\bar n$ transitions are calculated by integrating over the
corresponding operators in the extra dimensions. This model provides
further motivation for the future planned $n\bar n$ search
experiment at the ESS~\cite{Addazi:2020nlz}, see also
Refs.~\cite{Girmohanta:2019cjm,Nussinov:2020wri}.  Further, we discuss an
analysis of the effects of $n\bar n$ transitions in neutron stars
 \cite {Goldman:2024yoh}. Each $n\bar
n$ transition leads to $\Delta B=-2$ annihilation, releasing $\sim
2m_n$ energy. We calculate the effect on the luminosity of the neutron
star and find that it is negligibly small.  We also calculate the
effects on the rotation of neutron stars and on the periods of binary
pulsars, again finding these to be negligibly small. 

\newpage
\subsection{Binary pulsars, baryogenesis, and Majorana GeV Dark Matter}
\label{sec:binary}
\emph{Adrian Thompson}

Rare processes in the laboratory and within astrophysical environments can be highly sensitive probes of BNV interactions across a multitude of search strategies~\cite{Barrow:2022gsu}. Motivated by baryogenesis, some solutions to the dark matter and baryon number coincidence puzzle predict BNV operators that involve a GeV-scale dark sector field. These operators would ordinarily give rise to proton decay, but can be forbidden kinematically by the mass of the dark sector particle $\psi$ in the final state. However, the dense nuclear matter in neutron stars lifts the center-of-mass frame energy of baryons~\cite{Walecka:1974qa,Dexheimer:2008ax} such that baryon decays, e.g., $\mathcal{B} \to \psi \gamma$ to these dark sector particles can become kinematically accessible inside the star.

In Ref.~\cite{Allahverdi:2024ygd} we demonstrate the sensitivity of neutron star observables to constrain BNV by considering a minimal extension of the SM involving a TeV-mass color-charged scalar mediator and a GeV-scale Majorana fermion $\psi$~\cite{Dev:2015uca,Allahverdi:2017edd}.  The low energy effective theory leads to a new operator in the chiral perturbation theory Lagrangian,
\begin{equation}
\mathcal{L}_{\chi \rm PT} \supset \beta \text{Tr} [ \hat{C}^R u^\dagger B_R \psi u],
\end{equation}
where $B_R$ is the meson octet, $u = e^{i \Phi / f_\pi}$ for the meson octet $\Phi$, $\beta$ is related to the proton-to-vacuum matrix element, $\hat{C}^R$ is a $3\times3$ spurion matrix, and $B_R \psi$ is a shorthand for $b_{R}^{\dagger}\left[-i\sigma^2\right]\psi_{R}^* = \bar{b} P_L \psi^c$ where $b_R$ are the right-handed baryon fields.
Expanding this trace and additionally utilizing the baryon magnetic dipole moment, we find that an effective $\Delta B = 2$ mass loss process can take place in neutron stars via $n \to \psi \gamma$ decays (inspired by Refs.~\cite{Fajfer:2020tqf,Alonso-Alvarez:2021qfd,McKeen:2018xwc,McKeen:2020oyr,McKeen:2021jbh,Ema:2024wqr} for example), and the subsequent scattering $\psi n \to \pi^- K^+$, similar to ``induced nucleon decay'' processes whose phenomenology has been explored in other contexts~\cite{Davoudiasl:2010am,Davoudiasl:2015mcm,Davoudiasl:2011fj,Huang:2013xfa,Berger:2023ccd,Fox:2024kda}. This two-step mass loss process faces stringent constraints on the model parameter space. These constraints follow from the observed rate of orbital period decay in several binary systems~\cite{Berryman:2023rmh, Berryman:2022zic,Zakeri:2023xyj, Gardner:2023wyl}. Crucially, the Majorana nature of $\psi$ and the induced nucleon decay channel does not permit a sizable population of $\psi$ to build up in the star. In that case, the impact on the neutron star equation of state (EoS) and mass-radii relations or neutron star heating could be considered instead, like in the case of $n$--$n^\prime$ mixing~\cite{Silagadze:2023pxy,Berezhiani:2021src,Berezhiani:2020zck} and in other $\mathcal{B} \to \psi \gamma$ scenarios where the dark sector states do not completely annihilate~\cite{McKeen:2018xwc,McKeen:2020oyr,McKeen:2021jbh,Ema:2024wqr}.

However, the constraints we derive from pulsar timing when $\psi$ annihilation does take place still exclude much of the parameter space for $m_\psi \lesssim 1.4$ GeV at a comparable level to those without $\psi$ annihilation. These limits become much stronger in the model under consideration due to the possibility of $\Lambda \rightarrow \gamma \psi$ decays at the tree level, if the neutron star EoS is hyperonic. Meanwhile, the neutron--$\psi$ mixings that drive $n \to \psi \gamma$ are generated at one loop and have various amounts of CKM matrix element suppression. We compare these constraints, for several model EoS, with ongoing and future collider experiments~\cite{CDF:2012obh,Khachatryan:2014uma,Undleeb:2017oor,Aad:2014wza,Sirunyan:2018gka,Kalsi:2024hfe}, $n$--$\bar{n}$ oscillation experiments~\cite{Rao:1982gt,Phillips:2014fgb,SNO:2017pha,Super-Kamiokande:2020bov,Backman:2022szk,MicroBooNE:2023dci}, and dinucleon decay searches ($pp \to K^+ K^+$~\cite{Super-Kamiokande:2014hie,Litos:2010zra,Super-Kamiokande:2015pys}) across the landscape of couplings in the model that give rise to different flavor structure. We also motivate a dark matter direct detection search from $\psi n \to \pi^- K^+$ induced nucleon decay, but find that typical sensitivity to this channel is not as powerful as the bounds from binary pulsars and $pp \to K^+ K^+$ unless the latter constraints can be evaded in a model-dependent scenario; in this case, 21-cm limits on decaying dark matter severely restrict the parameter space, see, e.g., Ref.~\cite{Sun:2023acy}. We find that the binary pulsar bounds on couplings can be significantly tighter for specific flavor combinations, while other flavor coupling combinations have more CKM suppression in the $n \to \psi \gamma$ lifetimes that drive the baryon loss, leading to a higher degree of complementarity with terrestrial search strategies.

\newpage
\subsection{Probing new mechanisms of baryon number violation through pulsar-timing constraints in neutron stars}
\label{sec:neutron_star}
\emph{Susan Gardner}
 
BNV is a predicted outcome in many BSM models. 
Here we enlarge the definition of ``BNV'' to include 
{\it apparent} BNV in which a neutron disappears to a final state with 
non-SM particles and no baryons~\cite{Berryman:2022zic}. 
Neutron stars offer an enormous
baryon reservoir  in which such processes can be probed, with some $10^{57}$ neutrons expected in a 
$\sim 1.4 M_\odot$ 
neutron star~\cite{Baym:1975mf}---a number some $10^{23}$
the volume of any terrestrial proton-decay experiment~\cite{Berryman:2022zic}. 
This and the existence of 
precisely determined energy-loss constraints in neutron stars, particularly through pulsar-timing observations in binary-pulsar systems~\cite{Weisberg:2016jye}, 
limit not only the total baryon loss rate across the star
but also the parameters of the particle physics models that can produce such loss~\cite{Berryman:2023rmh,Allahverdi:2024ygd}. 

BNV is one of the Sakharov conditions for baryogenesis in the early Universe~\cite{Sakharov:1967dj},
but in low-scale baryogenesis models with ``hidden'' sectors that explain both dark matter and the cosmic 
baryon asymmetry, as in, e.g.,  Refs.~\cite{Davoudiasl:2010am,Allahverdi:2017edd,Elor:2018twp}, the noted
conditions can be sufficient, rather than necessary. 
A common feature of such scenarios is a generalized definition of baryon
number, across visible and hidden 
sectors~\cite{Davoudiasl:2012uw}. 
Even in the event that the total such 
baryon number is zero, a cosmic baryon asymmetry arises from generating net
baryon number in the visible sector, making processes that link baryons and hidden-sector baryons 
necessary. 
These models can also be ``EDM safe,'' in that they can yield 
baryogenesis even in the absence of new sources of $CP$ violation. 
Here, in collaboration with Jeffrey Berryman and Mohammadreza
Zakeri, we compute baryon decays to particles of a hidden sector,
in the dense nuclear medium found at the core of a neutron star~\cite{Berryman:2023rmh}. 
Our computational framework employs the techniques of
relativistic, nuclear 
mean-field theory~\cite{Walecka:1974qa,Serot:1984ey} and uses different neutron-star equations of state~\cite{Dexheimer:2008ax,CompOSECoreTeam:2022ddl}, but the detailed limits 
also depend on the new-physics model we employ.

An intriguing anomaly of decades-long standing lies in determinations of the neutron lifetime, in that its measured value 
depends on whether a surviving neutron or its decay products are counted~\cite{Wietfeldt:2011suo}. 
The lifetime difference is about a 1\% effect. 
Perhaps the neutron decays to a non-SM final-state, explaining why counting ``living'' neutrons yields a shorter lifetime~\cite{Fornal:2018eol,Berezhiani:2018udo}, though severe constraints
on this arise from the measured 
$V-A$ structure
of the SM weak currents 
in neutron decay~\cite{Czarnecki:2018okw}
and the observed neutron star masses~\cite{Baym:2018ljz,McKeen:2018xwc,Motta:2018rxp}. 
Yet, even so, the rate of baryon disappearance to exotic final states may grossly exceed that of neutron decay with BNV to SM final states. 
Intriguingly, a model with a baryon-number-carrying scalar that produces $n\to \chi \gamma$~\cite{Fornal:2018eol}, where $\chi$ is a dark baryon, 
also operates as a dynamical ingredient in ``$B$-mesogenesis''---a low-scale, dark co-genesis 
model~\cite{Elor:2018twp,Alonso-Alvarez:2021qfd}. 
There, out-of-equilibrium decays
of a scalar to a $B\bar B$ meson pair, with $CP$ violation, 
yields $B$ (or $\bar B$) decay to
${\cal B} \bar \chi$ (${{\bar{\cal B}}} \chi$); and a rate excess to 
the channel with a SM baryon ${\cal B}$ generates the cosmic baryon asymmetry. 
To do this, a dark-sector interaction 
must be chosen to mediate $\chi$ decay, 
to avoid washout and to generate a dark-matter candidate---and different choices and neutron-star scenarios~\cite{Gardner:2023wyl} are possible. 
At very low energies, these models lead to $n$--$\chi$ mixing at some strength.
If dark-sector particles do not accumulate 
in the star, its 
structure is fixed by its central energy density $\varepsilon_c$ as per the solution to the Tolman--Oppenheimer--Volkoff equations for a star with a fixed equation of state.
Supposing a quasi-equilibrium condition $\Gamma_{\rm BNV} \ll \Gamma_{\rm weak}$, where the latter characterizes the Urca (neutrino emission) rates~\cite{Gamow:1941gis}, this implies that as $\varepsilon_C$ changes from $n$ disappearance, the structure
of the star is still fixed by 
SM physics~\cite{Berryman:2022zic}. 
Thus, given a rate of change of ${\cal B}$, we can predict changes in the macroscopic parameters of the star and limit 
microscopic (dark-decay) models using relativistic mean-field theory. 
Using a suitable hidden-sector choice, we 
have followed this path to find severe 
constraints on $n$--$\chi$ mixing, noting that different choices of
binary-pulsar systems help us to sample the possible parameter
space, and thus to 
exclude this mechanism as
a fractional explanation of the neutron 
lifetime anomaly---and also to constrain 
the flavor structure of the noted co-genesis model~\cite{Berryman:2023rmh}.

\newpage

\section{Acknowledgments}

We thank the Institute for Nuclear Theory at the University of Washington for hosting this workshop INT-25-91W, supported by the INT's U.S.\ Department of Energy grant No.\ DE-FG02-00ER41132 and the Swiss National Science Foundation under Project No.\  PCEFP2\_181117.
The work of L.J.B.\ was supported by the U.S.\ Department of Energy, Office of Science, Office of Nuclear Physics (contract DE-AC05-00OR22725). 
%
A.C.\ and M.H.\ acknowledge support by the  Swiss National Science Foundation (Project Nos.~PP00P21\_76884 and  TMCG-2\_213690).
S.S. is supported by the US National Science Foundation under Award PHY-2412963.
This manuscript has been authored by Fermi Forward Discovery Group, LLC under Contract No.\ 89243024CSC000002 with the U.S.\ Department of Energy, Office of Science, Office of High Energy Physics.
Y.A.~thanks Yoshinobu Kuramashi, Eigo Shintani and Ryutaro Tsuji (through PACS collaboration) for their collaboration on the project, which is supported in part by the MEXT Program for Promoting Researches on the Supercomputer Fugaku: Large-scale lattice QCD simulation and development of AI technology, JPMXP1020230409.
A.B.B.\ acknowledges funding from the Spanish ``Agencia Estatal de Investigación,'' MICIN/AEI/10.13039/501100011033, and from the grant CIACIF/2021/061. He also thanks Juan Herrero García, Arcadi Santamaria, John Gargalionis, and Michael Schmidt for their collaboration on the work presented here.
The work of Z.B.\ was partially supported by the research grant number 2022E2J4RK ``PANTHEON: Perspectives in Astroparticle and
Neutrino THEory with Old and New messengers'' under the program PRIN 2022 (Mission 4, Component 1, CUP I53D23001110006) funded by the Italian Ministero dell’Universit\`a e della Ricerca (MUR) and by the European Union – Next Generation EU. 
N.F.C.~thanks the Hyper-Kamiokande Collaboration and INFN for their support.
S.G.~thanks the organizers for the opportunity to contribute to this document, despite her unexpected inability to attend, as well as 
Jeffrey Berryman and Mohammadreza Zakeri for their collaboration on the work presented here, and she acknowledges
support from the U.S.\ Department of Energy, Office of Science, Office
of Nuclear Physics under contract DE-FG02-96ER40989. 
J.H.~thanks Diana Sokhashvili and Anil Thapa for collaboration on some of the work presented here and acknowledges support by the U.S.~Department of Energy under Grant No.\ DE-SC0007974 and a 4-VA at UVA Collaborative Research Grant.
C.J.~thanks Yuxiang Hu, Wanlei Guo, LiangJian Wen, and JUNO collaboration for their collaboration in the research presented here.
A.A.P.'s research was supported in part by the US Department of Energy grant DE-SC0024357. He thanks Martin Beneke and Gael Finauri for the collaboration on the work presented here.
The research of R.S.\ was supported in part by the U.S.\ National Science Foundation grant NSF-PHY-22-10533. He thanks S.~Girmohanta, I.~Goldman,
R.~Mohapatra, and S.~Nussinov for collaboration on the results presented here.
A.T.~thanks Mohammadreza Zakeri and Rouzbeh Allahverdi for the collaboration on the work presented here and acknowledges support in part by the DOE grant DE-SC0010143.
The work of U.v.K.\ was supported in part by the US Department of Energy, Office of Science, Office of Nuclear Physics, under award DE-FG02-04ER41338. He acknowledges the essential contribution to the work reported here from his collaborators Femke Oosterhof, Jordy de Vries, Bingwei Long, and Rob Timmermans.
J.W.~thanks Lund University and ESS for their support of his travel and D.~Milstead for Fig.~\ref{sensitivity}.

\newpage

\section*{References}

\bibliographystyle{apsrev4-1_mod_iop}
\bibliography{BNV}

\end{document}